\documentclass[journal]{IEEEtran}
\usepackage{caption}
\usepackage{cite}
\ifCLASSINFOpdf
  \usepackage[pdftex]{graphicx}
  \graphicspath{{./FigurePaperFinal/}}
\else
  \usepackage[dvips]{graphicx}
  \graphicspath{{./FigurePaperFinal/}}
\fi
\usepackage[cmex10]{amsmath}
\usepackage{amsfonts}
\newcommand{\e}[1]{{\mathbb E}\left[ #1 \right]}
\usepackage{array}
\usepackage{url}
\usepackage{xfrac }
\usepackage{color,soul}

\begin{document}
\bstctlcite{IEEEexample:BSTcontrol}
\newtheorem{definition}{\textbf{Assumption}}
\title{Recursive Discrete-time Models for Continuous Time Systems under Band-limited Assumptions}
\author{Rishi~Relan,
        Johan~Schoukens,~\IEEEmembership{Fellow,~IEEE}%
\thanks{The  authors  are  with  the  ELEC  Department,  Vrije  Universiteit  Brussel (VUB),Brussels B-1050, Belgium (e-mail: Rishi.Relan@vub.ac.be, Johan.Schoukens@vub.ac.be). This work was supported in part by the Fund for Scientific Research (FWO-Vlaanderen), by the Flemish Government (Methusalem), the Belgian Government through the Inter university Poles of Attraction (IAP VII) Program, and by the ERC advanced grant Structured Nonlinear System Identification (SNLSID)} , under contract 320378.
}
\markboth{ArXiV preprint: The article is published in IEEE Trans. Instrum. Meas.Vol. 65, Issue: 3, March 2016, DOI: 10.1109/TIM.2015.2508279 }%
{Shell \MakeLowercase{\textit{et al.}}: Bare Demo of IEEEtran.cls for Journals}
\maketitle

\begin{abstract}
Discrete-time models are very convenient to simulate a nonlinear system on a computer. In order to build the discrete-time simulation models for the nonlinear feedback systems (which is a very important class of systems in many applications) described as $ y(t) =g_{1}(u(t),y(t))$, one has to solve at each time step a nonlinear algebraic loop for $y(t)$. If a delay is present in the loop i.e $ y(t) = g_{2}(u(t),y(t-1))$, fast recursive simulation models can be developed and the need to solve the nonlinear differential-algebraic (DAE) equations is removed. In this paper, we use the latter to model the nonlinear feedback system using recursive discrete-time models. Theoretical error bounds for such kind of approximated models are provided in the case of band-limited signals, furthermore a measurement methodology is proposed for quantifying and validating the output error bounds experimentally.  
\end{abstract}


\IEEEpeerreviewmaketitle

\section{\textbf{Introduction: Discrete-time Modeling}}

Modern measurement instruments make frequent use of advanced signal processing and control algorithms that are designed as well as implemented using discrete-time (non) linear models. Since most of the real-world systems evolve in continuous time, it should be carefully checked, if a discrete-time model can be used to describe such systems, and what errors might be created in the discretization step.  Especially under the band-limited assumption (signals have no power above a given maximum frequency,  for example measurements using anti-alias protection) this question becomes important. While for the linear systems, the error mechanism is well understood, it turns out that, it is not obvious how to quantify these errors for the nonlinear  systems. This problem is addressed in this paper. This paper analyses first the nature of the error, and using these insights, it is shown how the measurement procedures can be designed in order to keep the error below an user specified level by making a proper choice for the sampling frequency. 

This  is  also  an  important  and  relevant  problem for the instrumentation and measurement community since in many instruments nonlinear  post- or pre-processing is  done. Sensor linearization (e.g. for the relative humidity or temperature sensors) and nonlinear pre-compensation of actuators (Actuator nonlinearities can be static like friction, deadzone, saturation, and/or dynamic in nature like backlash and hysteresis due to the inaccuracy in the manufacturing of mechanical components and due to the nature of the physical laws \cite{selmic2003intelligent}) \cite{yu2008nonlinear, gubian2009study} are typical examples. In modern instruments pre-compensation and linearization must be done on-board in real-time, hence use of a sophisticated numeric solver at each iteration is not feasible due to real-time constraints. Therefore discrete-time models for continuous-time systems with bounded output errors must be developed to deal with such constraints. Moreover, the goal of many measurement procedures is  to build a discrete-time (nonlinear) simulation model for  various real world systems such as gas sensing systems, gas turbine engines, large-signal amplifiers, RF thermistors etc. \cite{Pardo1998, chiras2001nonlinear, mirri2004nonlinear, kazemipour2011nonlinear, baglio2011exploiting, lota2013nonlinear}. Discrete-time models for linear and non-linear dynamical systems can be developed under different assumptions of the measurement set-up, e.g. the zero-order hold (ZOH) and the band-limited measurement set-up. 

\subsection{Linear Systems}

 \subsubsection{Zero-order hold measurement set-up} 
\begin{figure}[!h]
 \centering
\captionsetup{justification=centering}
 \includegraphics[width=0.5\textwidth]{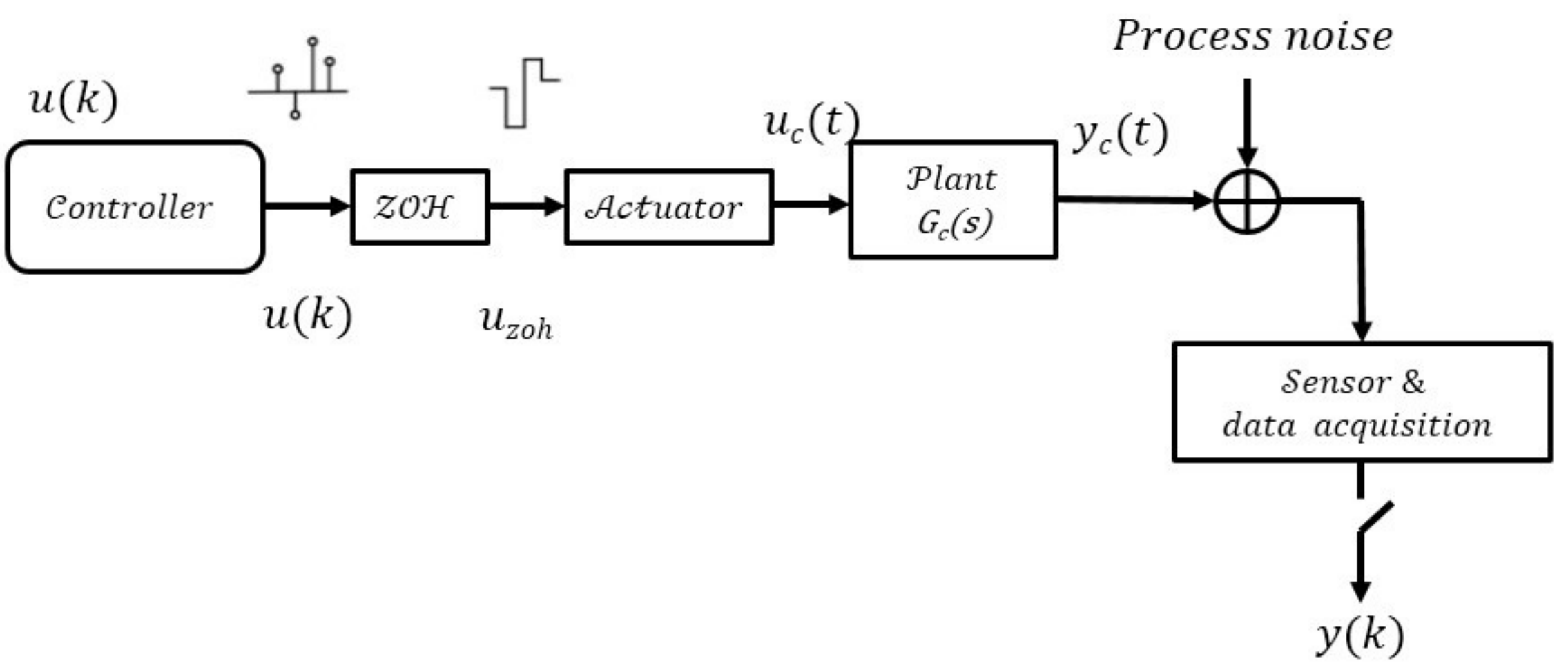}
\caption{Zero-order hold (ZOH) measurement set-up }
\label{ZOH}
\end{figure}

In the case of a zero-order hold (ZOH) measurement set-up, as shown in Fig.\ref{ZOH}, the linear-time invariant system \cite{ljung1998} can be described using the discrete-time representation: 

\begin{equation}
y(t) = \sum\limits_{k=1}^{\infty} g(k)u(t-k)
\label{eqn:LZOH}
\end{equation}

\subsubsection{Band-limited measurement set-up}
\begin{figure}[!ht]
 \centering
\captionsetup{justification=centering}
 \includegraphics[width=0.5\textwidth]{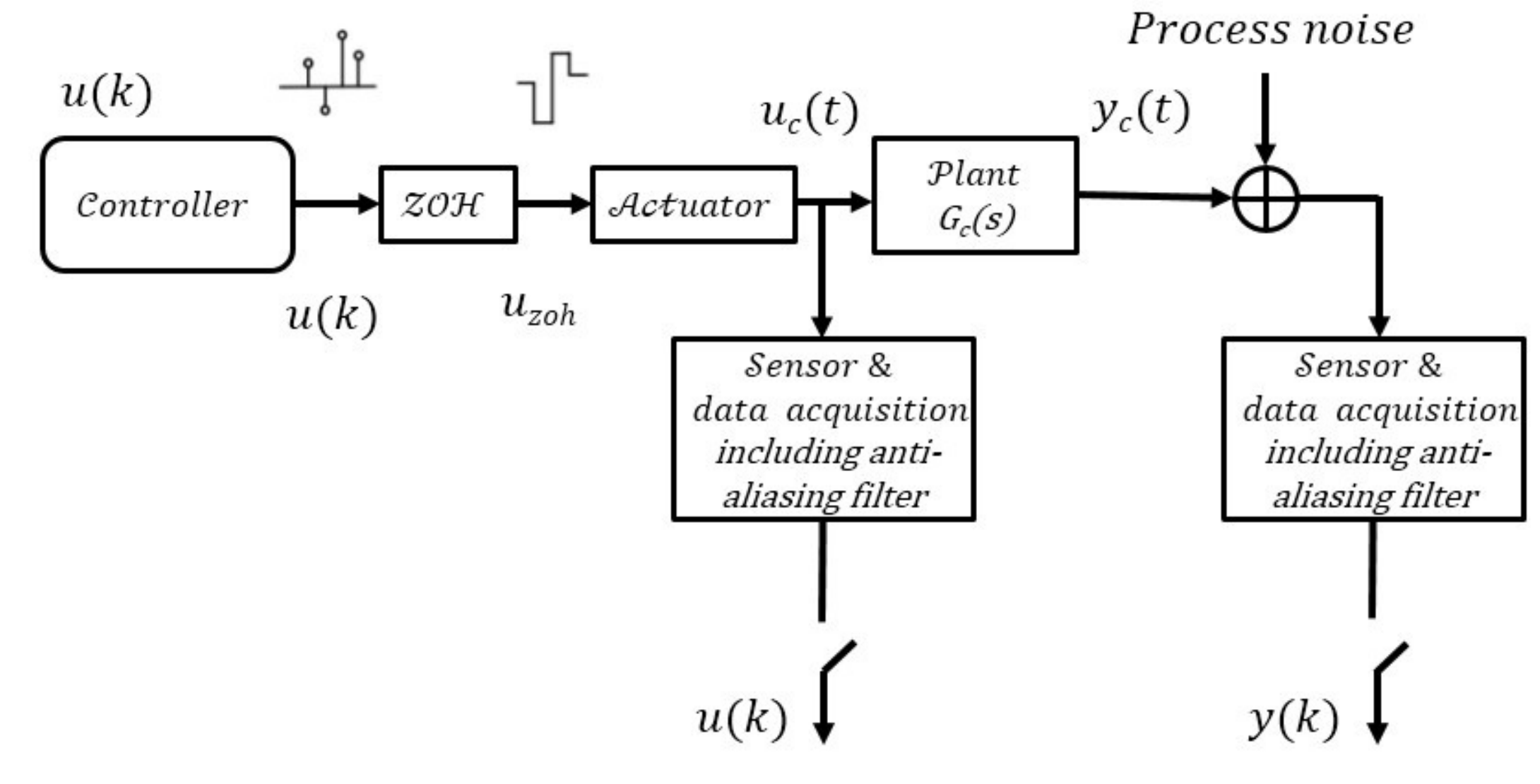}
\caption{Band-limited measurement set-up}
\label{BL}
\end{figure}

For the band-limited measurement setup, as shown in Fig.\ref{BL}, the discrete-time representation of the system can be written as

\begin{align}
y(t) &= \sum\limits_{k=0}^{\infty} g(k)u(t-k)\notag\\
&= g(0)u(t) + \sum\limits_{k=1}^{\infty} g(k)u(t-k)
\label{eqn:Bl}
\end{align}

It can be seen in (\ref{eqn:LZOH}) that it does not contain any direct term i.e $ g(0)$ term. This kind of representation is very popular in discrete-time control systems whereas the discrete-time representation (\ref{eqn:Bl}) contains the direct-term. This kind of model representation is very popular in the digital signal processing community and it is more appropriate for simulation as well as measurement applications.


\subsection{Nonlinear Systems}
\label{subsec:Nonlinear Systems}
One would reasonably expect similar results to hold for the nonlinear systems. However, the situation for the nonlinear case is more complex than for the linear systems. In the case of a zero-order hold (ZOH) measurement set-up, we would like to write the output of the system as 
\begin{equation}
y(t)=g_{1}(u^{t-1},y^{t-1})
\label{eqn:NonZOH}
\end{equation} where $u^{t} = [u(t),u(t-1),......u(1)]$ and $y^{t} = [y(t),y(t-1),......y(1)]$. In general this does not hold true for e.g. nonlinear feedback systems. In the case of nonlinear feedback systems (\ref{eqn:NonZOH}) becomes $y(t)=g_{1}(u^{t},y^{t})$ and one needs to solve nonlinear algebraic loops due to the presence of a direct-term. Also for the band-limited measurement set-up similar constraints exist. The questions which we would like to raise is whether we can approximate $y(t)=g_{1}(u^{t},y^{t})$ by $y(t)=g_{1}(u^{t-1},y^{t-1})$, and how the approximation error depends on the experimental conditions.

\begin{figure}[!h]
 \centering
\captionsetup{justification=centering}
 \includegraphics[width=0.45\textwidth]{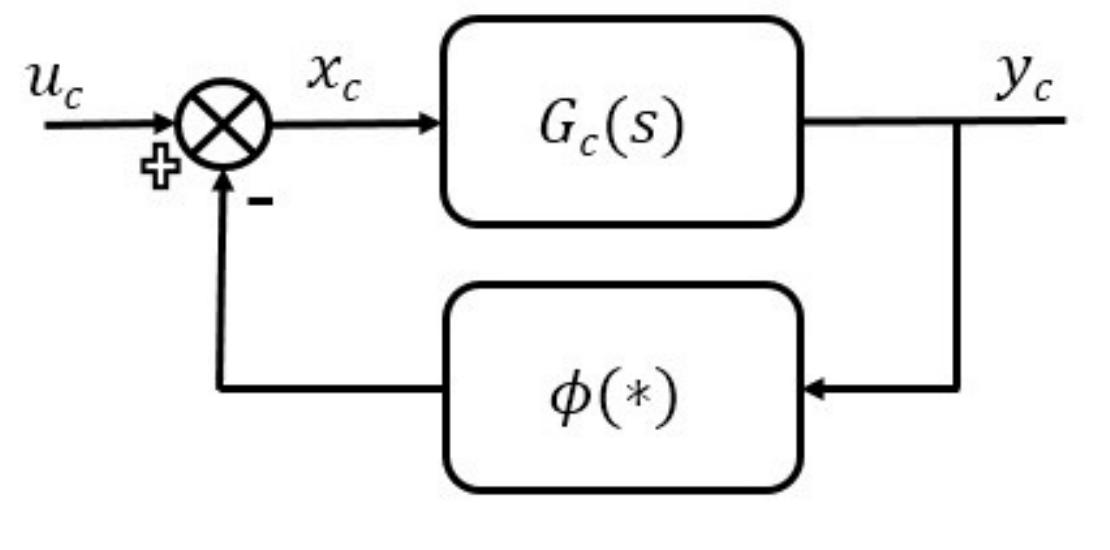}
\caption{Nonlinear feedback system: continuous time}
\label{NonFeedSys}
\end{figure}

Consider for example a nonlinear feedback system as shown in Fig.\ref{NonFeedSys}, where $ G_c(s) $ is a Laplace transfer function between the input signal $ \it{x}_{c}(t) $ and the output  signal $ \it{y}_{c}(t)$. $ \MakeUppercase{\phi(*)} $ is any memory-less, static non-linearity in the feedback loop. Many  electrical, electronic and physical  systems, e.g. oscillators \cite{Wambacq1998}, biomedical \cite{Marmarelis1991} and mechanical system \cite{Kerschen2006, Worden2001, Bucher1998}, contain in  an  implicit  manner,  a  nonlinear  feedback loop and can be described using the similar model structure.  

\begin{figure}[!h]
 \centering
\captionsetup{justification=centering}
 \includegraphics[width=0.45\textwidth]{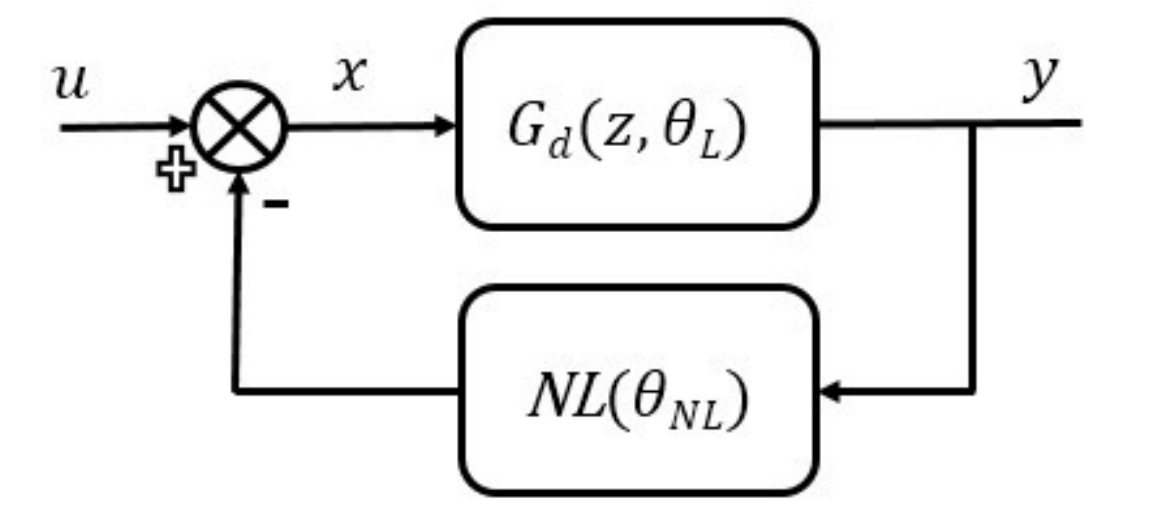}
\caption{Nonlinear feedback system: discrete-time}
\label{NonFeedSysD}
\end{figure}Fig.\ref{NonFeedSysD} shows a possible model structure for a discrete-time representation of the continuous-time system in Fig.\ref{NonFeedSys}. For a band-limited input signal $u_{c}(t)$ (see later for a precise definition), the linear system $\MakeUppercase{g}(s)$ can be approximated by a discrete-time model

\begin{equation}
y(t) = \sum\limits_{k=0}^{\infty} g(k)x(t-k),
\label{eqn:DM}
\end{equation} and $q^{-1}$(one sample delay operator) provided that the sampling frequency $f_{s}$ is sufficiently high such that the aliasing errors are acceptably small. For a given sampling period $\MakeUppercase{T}_{s}= \frac{1}{f_s}$, the discrete-time signals $u$ and $y$ can be represented as

\begin{equation}
u(k)=u_{c}(kT_{s})\hspace{0.01\textwidth};
\hspace{0.02\textwidth}
y(k)=y_{c}(kT_{s}).
\end{equation}

The output of the discrete-time model described by Fig.\ref{NonFeedSysD} can be written as 

\begin{equation}
y(t) = g(q,\theta)(u(t)-\phi(y(t))).
\label{eqn:DAENL}
\end{equation} This is an example of a nonlinear algebraic loop \cite{Markusson2001}, which implies that a set of nonlinear differential-algebraic equations (DAE) should be solved at each time step in order to calculate the model output. In the control engineering community, many approaches dealing with discrete-time representation of continuous-time systems implicitly assume that the direct term is equal to zero, in order to deal with the problem. In principle, this problem (of continuous time system modelling and simulation) can be tackled by utilising dedicated numeric (integration) solvers. The main disadvantage of using a numerical integration solver is, that it can have multiple solutions. In fact, it is even possible that no solution exists \cite{Paduart2007}. Moreover, it is a very time consuming approach as well as the robustness of the obtained solution can not be guaranteed \cite{Cellier1991, Cellier2006} therefore these models are not well suited for real-time applications. Interested readers are referred to the Section \ref{SampData} for a brief introduction to the literature dealing with the issues of sampled data models within the control system community. The identification of block-oriented nonlinear feedback models received considerably less attention and still is in its infancy. To the best of authors knowledge, this issue of (how to avoid) nonlinear algebraic loop has not been tackled before in the instrumentation and measurement community. 

The authors in \cite{schoukens2008identification,gomme2009baseband,gomme2009time} used block-oriented nonlinear feedback structure to model a microwave crystal detector RF applications, but it turns out that the nonlinear algebraic loop created convergence problems for larger inputs. In order to avoid the nonlinear algebraic loops while developing the discrete-time  nonlinear linear fractional representation (LFR)  model (representation of nonlinear feedback systems), the authors in \cite{LaurentImtc2012, vanbeylen2013improved, LVanImproved2013, AnneImtc2013, AnneECC2014} assumed that the one tab delay is present implicitly in the loop or in other words the direct feed-through term is $0$. The authors did not check the validity (neither theoretically nor experimentally) of their assumption. This assumption fails under the band-limited measurement conditions as explained above, whereas \cite{borin2000elimination} proposed a solution by means of geometrical transformation of the nonlinearities and algebraic transformation of the time-dependent equations in order to deal with algebraic loops in nonlinear acoustic systems. This approach may not be optimal for fast recursive models intended for real-time scenarios.  

Hence, the main idea in this paper is to show under which experimental conditions and constraints we can develop a discrete-time recursive simulation model. To do this in a simple (similar) way, we can impose one sample delay for the linear block or, equivalently $g(0)$ in (\ref{eqn:DM}) will be set to zero. Taking into account the imposed delay, we will obtain the following model equation

\begin{align}
y(t) &= \sum\limits_{k=1}^{\infty} g(k)x(t-k)\notag\\
x(t)&=u(t)-y(t)\notag\\
y(t) &= \sum\limits_{k=1}^{\infty} g(k)(u(t-k)-\phi(y(t-k)) 
\label{eqn:DTNL}
\end{align} Under the band-limited assumption this discrete-time representation is a recursive in nature. Hence, we can develop the recursive discrete-time simulation models by forcing the direct-term of the identified model equal to 0, i.e. by introducing explicitly a delay in to the loop. In order to do this, some associated questions need to be answered:

\begin{itemize}
\item How to quantify the errors associated with the approximated models ?
\item What are the different factors/parameters which can influence the errors ?
\item How can we keep the error in the approximated model small enough by choosing the appropriate experimental conditions ?
\end{itemize} In order to answer these questions, in this paper, we propose a measurement approach to analyse and bound the output error of the developed discrete-time model for band-limited measurements. The rest of the paper is organized as follows. Section II gives an overview about the identification of the sampled-data models. Section III formalizes the problem statement and provides a comprehensive theoretical analysis of the errors associated with approximated linear discrete-time models with direct-term equal to 0. Section IV gives an overview of an experimental investigation performed in order to validate the obtained theoretical bounds on the error qualitatively both for linear as well as nonlinear (nonlinear feedback) dynamical systems and Conclusions are formulated in Section V.

\section{\textbf{Sampled data Models: general remarks}}
\label{SampData}

Identification  of  continuous-time  systems  from sampled data \cite{SinhaRao1991} is  a problem  of  considerable  importance in the control system community. These discrete-time representations of the continuous time system can be developed under a zero-order hold (ZOH) or band-limited (BL) assumption of the inter-sample behaviour \cite{RikJohan2012}. Exact discrete-time models of a continuous time systems at the sampling instances can be obtained for the linear systems by assuming a zero-order hold (ZOH) input \cite{Astrom2011,Middleton1990,Feuer1996}. 
This argument may, however, lead to a false sense of security when using sampled data as the pole-zeros patterns of the discrete-time systems may not be similar due to the presence of the extra zeros called, \textit{sampling zeros}, in the associated discrete-time transfer function, which is the consequence of the sampling process. These zeros have no counterpart in the underlying deterministic continuous-time model \cite{Astrom1984}.  Further in-depth information about the sampled data models for linear (Non-linear) deterministic (Stochastic) system with different sample and hold characterizations can be found in \cite{HAGIWARA1993, Eduardo2011,Wahlberg1990, Yuz2005,Graham2008,YuzJuan2008,Goodwin2013, Yuz2014} and the references mentioned therein. For the purpose of this study, this is not so important, as here we focus on the input-output behaviour of the underlying system at the discrete-time samples. Most of the previous work assumes that the data is gathered under the zero-order hold assumption.

In these ZOH models, no direct term is present for the linear system, see also (\ref{eqn:LZOH}). As explained earlier, most of these assumptions do not hold true for complex nonlinear dynamical systems, including networked dynamical systems, as the signals in the loop are no longer ZOH. Hence, it is important to consider discrete-time models under band-limited measurement assumptions. Even under band-limited assumptions, an arbitrary good discrete-time representation of the underlying continuous-time system can be retrieved. However, in that case, the direct term will be different from zero. The main emphasis of this study is to analyze the impact of explicitly setting this term to zero.
The resulting error will not only depend on the signal properties (Band-limited in this study), but also on the relative degree of the underlying continuous-time system. Indeed as the sampling rate increases, a continuous-time system with relative degree $d \geq 2$ behaves as an $d^{th}$ order integrator $\frac{1}{s^d}$ beyond Nyquist frequency. Therefore, the impulse response will depart very slowly from zero. The direct term in the discrete-time representation of such continuous-time models will be small, and will diminish to zero as the sampling frequency is increased. In the next section, under the band-limited measurement assumptions, a detailed analysis of the error associated with the discrete-time models with direct term equal to $0$ is provided. 
 
\section{\textbf{Theoretical Analysis}}
\label{Theory}

\subsection{Proposed Methodology}

\begin{figure}[!h]
 \centering
\captionsetup{justification=centering}
 \includegraphics[width=0.5\textwidth]{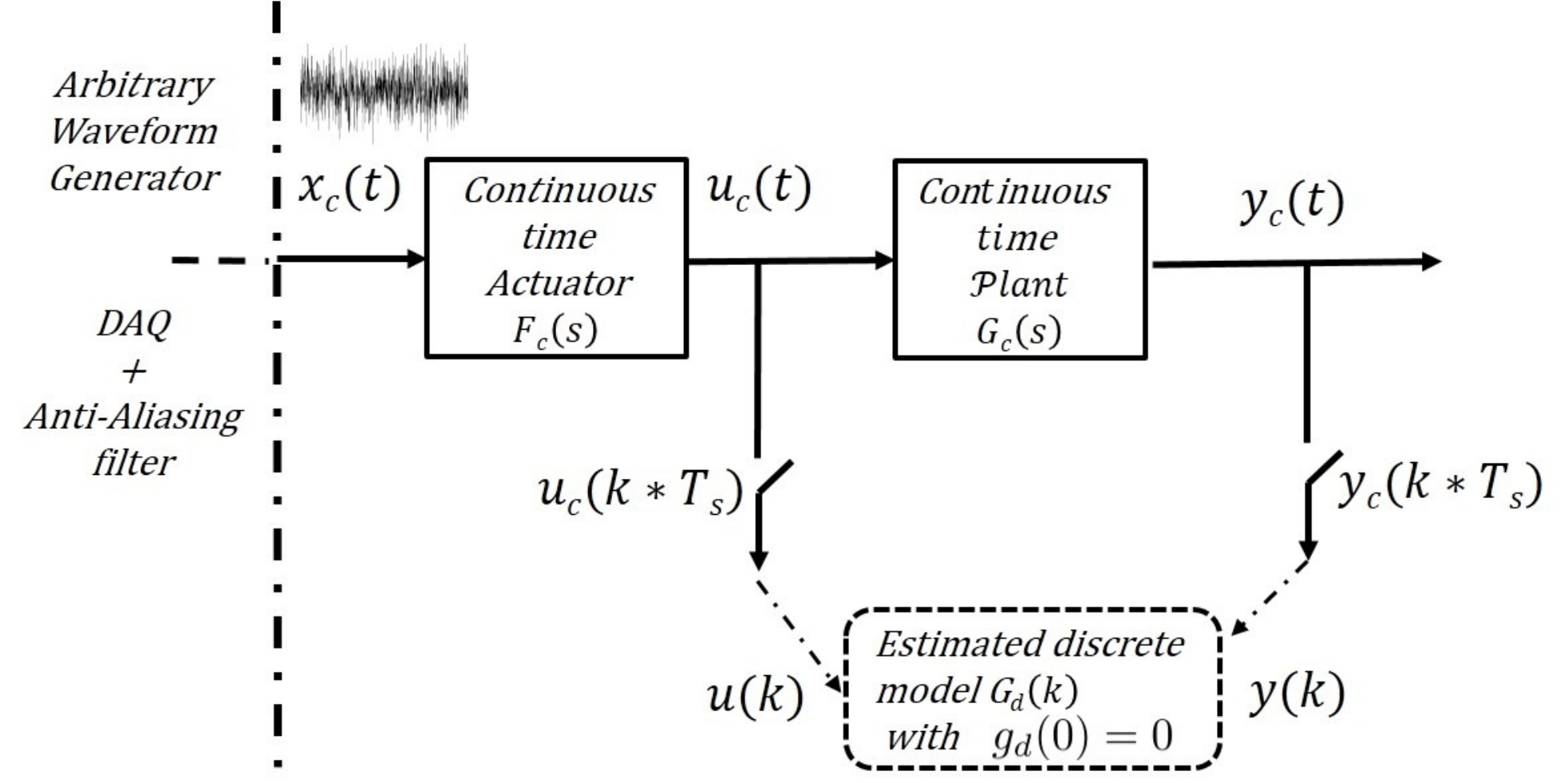}
\caption{Proposed methodology for the error quantification}
\label{ProAnalysis}
\end{figure}

In this section, we provide a thorough theoretical analysis based on the preliminary experimental investigations performed by \cite{Rishi2015I2MTC} using the measurement methodology as shown in Fig.\ref{ProAnalysis}. In this section under the band-limited measurement set-up assumptions, error bounds for the approximated linear discrete-time models with the direct-term set equal to 0 will be provided.

The aim in this study is to identify the discrete-time model $G_{d}(k)$, with direct-term $g_{d}(0)$ forced equal to 0, from the sampled measurements $u(kT_{s})$, $y(kT_{s})$ of the continuous-time plant $G_{c}(s)$, with input signal $u_{c}$ and output signal $y_{c}$, under band-limited measurement conditions. $T_{s}$ is the sampling period. $u_{c}$ can be an output from an actuator or a generator filter $F_{c}(s)$ which in turns can be excited by an arbitrary signal, e.g. white noise, random-phase multisines or any zero-order hold (ZOH). 

The main aspects which can influence the magnitude of the error in the output signal of the identified discrete-time model are:

\begin{itemize}
\item Is it possible to make an accurate prediction about the future input $u_{c}$ given its past sampled values i.e. $\hat{u}(t|t-1)$ ?
\item How much is the error $e(t)$ that would be introduced in case we can not make a perfect or accurate enough prediction of the input $u(t)$ = $\hat{u}(t) + e(t)$ ? 
\item What is the influence of the error in the one-step ahead prediction $\hat{u}(t|t-1)$ and of the direct $g(0)u(t) $ term on the final output signal of the discrete-time model i.e. $\hat{y}(t)$ in (\ref{eqn:Bl}) ?
\end{itemize} The reasoning mentioned above holds equally true in the case of block-oriented nonlinear feedback models as discussed in \ref{subsec:Nonlinear Systems}, because there are band-limited signals in the loop, hence it is enough for one system to have a delay to break the nonlinear algebraic loop. Hence, the first step in the analysis is to quantify the error in the one-step ahead prediction of the input signal $u(t)$ for a band-limited input signal. Next we analyse the effect on the system's output.


\subsection{Error Analysis}

In  this section, first a brief introduction to the band-limited signals and processes is given. Thereafter, to address the problem, the following steps mentioned below will be taken. 

\begin{enumerate}

\item quantification of the one-step ahead prediction error in the case of a perfectly BL signal $u(t)$;
\item quantification of the one-step ahead prediction error in the case of an actual BL signal $u(t)$, e.g., filtered ZOH signal;
\item quantification of the error in the final output signal $y(t)$. 
\end{enumerate}

\paragraph*{Band-limited signals and processes}

\begin{figure}[!h]
\centering
\captionsetup{justification=centering}
 \includegraphics[width=0.5\textwidth]{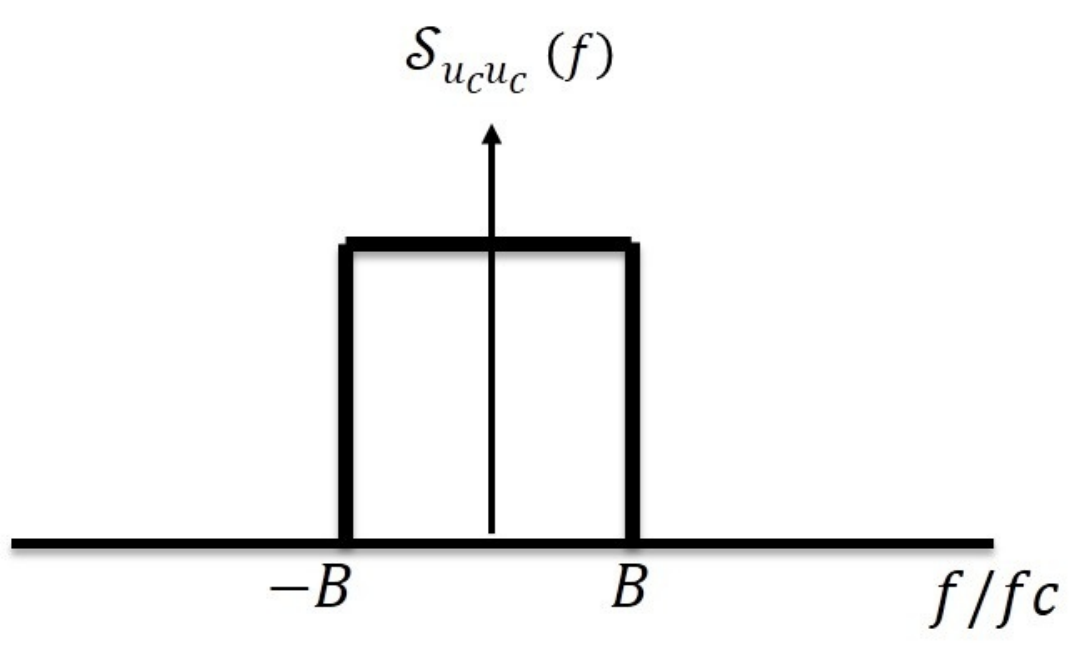}
\caption{Power spectrum of an ideal Bandlimited signal}
\label{BLBook}
\end{figure}

A signal is said to be BL (perfectly) if the amplitude of its spectrum goes to zero for all frequencies beyond some threshold called the cut-off frequency i.e., $U(j\omega) = 0$ for $|{\omega}| > \omega_B$. A wide sense stationary (WSS) random process is termed BL if its power spectral density (PSD) is BL, i.e., $S_{u_cu_c}(j\omega) = 0$ for $|{\omega}| > \omega_B$ is zero for frequencies outside some finite band. The power spectrum of a perfectly BL signal is shown in Fig. \ref{BLBook}. The next section describes briefly the theoretical aspects related to the prediction of perfectly BL signal and the associated error bounds with the prediction.

\subsection{One-Step Ahead Prediction Of A Band-limited $u(t)$}

\subsubsection{Error quantification:Perfectly band-limited signal}

An interesting problem in linear-prediction  theory is the  following: Let  $u_c(t)$ be a real continuous-time signal,  band-limited  to  the region $|{\omega}| \leq \omega_B$ or in other words consider a stationary stochastic process $u_c(t)$ with power spectrum 

\begin{equation}
S(\omega)= 0 \hspace{0.02\textwidth} for \hspace{0.02\textwidth}|\omega| > \omega_B=\frac{\pi}{T}
\end{equation} What is the smallest sampling frequency $f_s$, which will enable us to  predict  the present sample values $u_c(nT_s)$ based  on a finite  number  of  past  samples,   with an arbitrarily small  (pre-specified)  error,  and  with  predictor coefficients independent of the signal $u_c(t)$?  (Here $T_s$ represents the sampling period and $T_1$ is less than Nyquist rate $T$ i.e $T_1 < T $). For an arbitrary positive number $\epsilon > 0$, we can find a set of conditions such that

\begin{equation}
\hat{u}_c(t) = \sum\limits_{1}^{n} a_{n}u_c(t-nT_1),
\end{equation}

\begin{equation}
\e{\left\Vert\left(u_c(t)- \sum\limits_{1}^{n} a_{n}u_c(t-nT_1)\rm\right) \right\Vert^2}<\epsilon_1  .
\end{equation} For band-limited  processes  with  absolutely  continuous  spectral measure  i.e.,  processes  with  spectral  density,  Wainstein  and Zubakov \cite{Wainstein1971}  proved  that  if  the  sampling  rate  is  increased at  least  three  times  above  the  Nyquist  rate a band-limited process can be predicted with arbitrarily small error from its past samples using an universal formula for the predictor. A better result in this direction is \cite{Brown31972}, where a similar predictor is constructed when the samples are taken at twice the Nyquist rate. This sequence of predictors converges with exponential rate. However, it could be more difficult to find explicit coefficients for the predictor. These results  were further  improved  by Splettst{\"o}sser \cite{Splettstösser1982} in 1982, who showed that this kind of prediction  is possible even with  the sampling  frequency  equal   to 1.5 times the Nyquist  frequency.

Brown \cite{Brown31972} and Splettst{\"o}sser \cite{Splettstösser1982} have  also observed  that it is theoretically  possible to  predict  the  samples of $u_c(t)$  in  the above manner, as  long as the sampling  frequency is larger  than the Nyquist rate by any arbitrarily small amount $\epsilon_2 > 0 $. This observation has also been made by Papoulis \cite{Papoulis1985} who has given a different proof showing that the greatest lower bound of the prediction error is zero. 

\begin{equation}
\mathop{G.L.B.E}_{a_n}\left\{\e{\left\Vert\left(u_c(t)- \sum\limits_{1}^{\infty} a_{n}u_c(t-nT_1)\rm\right) \right\Vert^2}\right\} = 0  .
\end{equation} Further references, and  proofs can be  found  in \cite{Marvasti1986},\cite{Brown1986}, and  the references contained  therein. The result presented in  \cite{Brown1986} and \cite{Papoulis1985} are in fact particular cases of \cite{Beutler1961}. This discussion concludes that a band-limited signal can be perfectly predicted from  its past values (samples) provided that the sampling frequency $f_s $ is larger  that the Nyquist rate by any arbitrarily small amount $ \epsilon > 0 $.

\subsubsection{Error quantification:Low-pass filtered signal} 
\begin{figure}[!h]
\centering
\captionsetup{justification=centering}
 \includegraphics[width=0.5\textwidth]{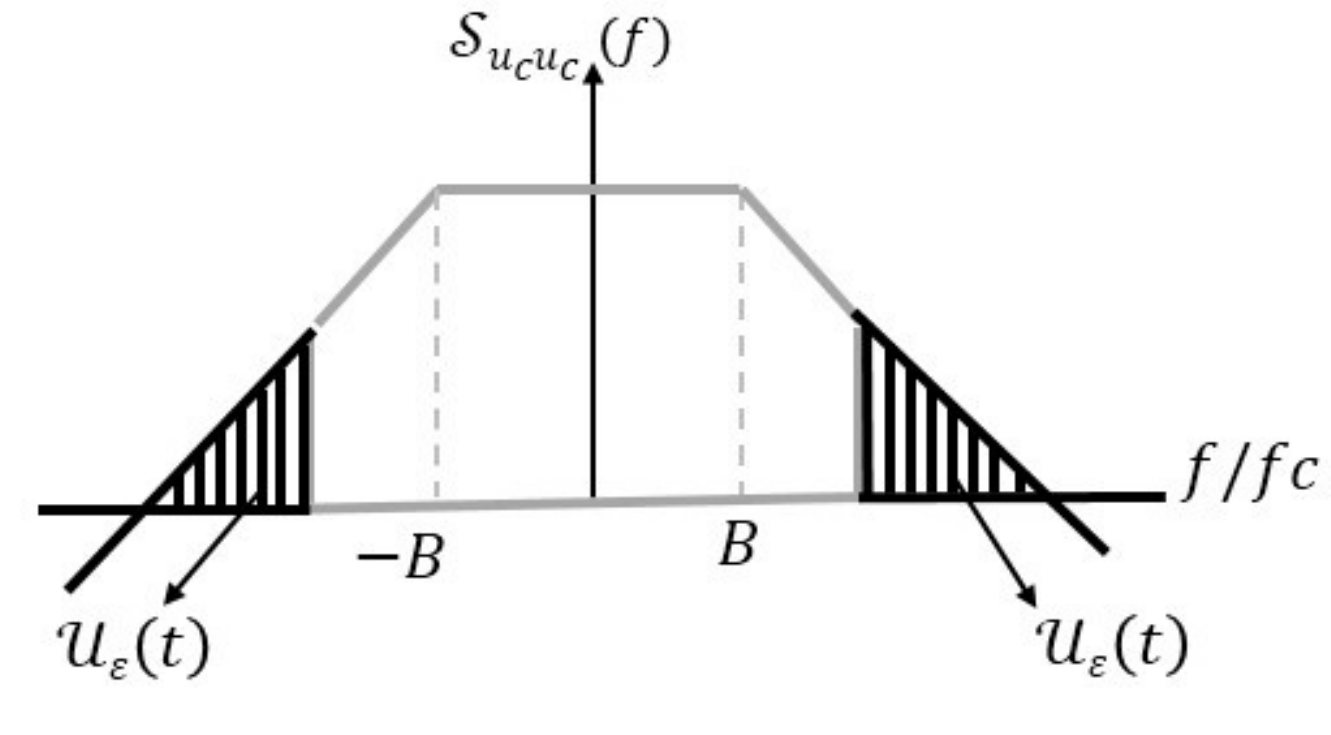}
\caption{Power spectrum of an actual bandlimited signal}
\label{BLReal}
\end{figure}

But in practice it is impossible to have a perfectly band-limited signal. In practice, a band-limited signal $u_c(t)$ can be considered to be made of two parts: $ u_{bl}(t)$ a part of the signal which can be perfectly predicted, and $u_\epsilon(t)$ which can not be predicted or remains unexplained as shown in Fig.\ref{BLReal} $$u_c(t) = u_{bl}(t) + u_\epsilon(t) $$

 Therefore the lower bound of the one-step ahead prediction error would be:

\begin{equation}
 \e{e(t)^2}\leq\e{u_{\epsilon}(t)^2}
\end{equation} Further in the discussion below a concise theoretical explanation is given to  quantify as well as to identify the factors associated with the error in a one-step ahead prediction of an actual  band-limited signal.  
 
\subsubsection{One-step ahead prediction of an actual band-limited $u(t)$}
\label{OneStepTheory}
\begin{figure}[!h]
\centering
\captionsetup{justification=centering}
 \includegraphics[width=0.5\textwidth]{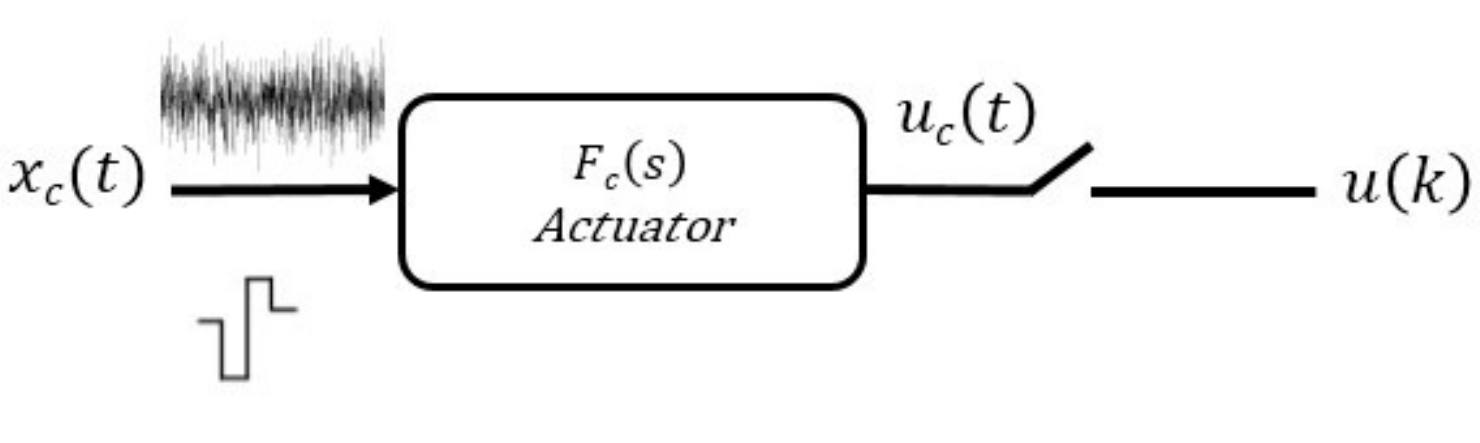}
\caption{Analysis of one-step ahead prediction}
\label{BLAna}
\end{figure}

Consider the case of a low pass filter $F_c(s)$ shown in Fig.\ref{BLAna}. for example, a Butterworth or a Chebyshev filter of order $n$, with cut-off frequency $f_c$, excited by white noise input signal. 

\begin{figure}[!h]
\centering
\captionsetup{justification=centering}
 \includegraphics[width=0.5\textwidth]{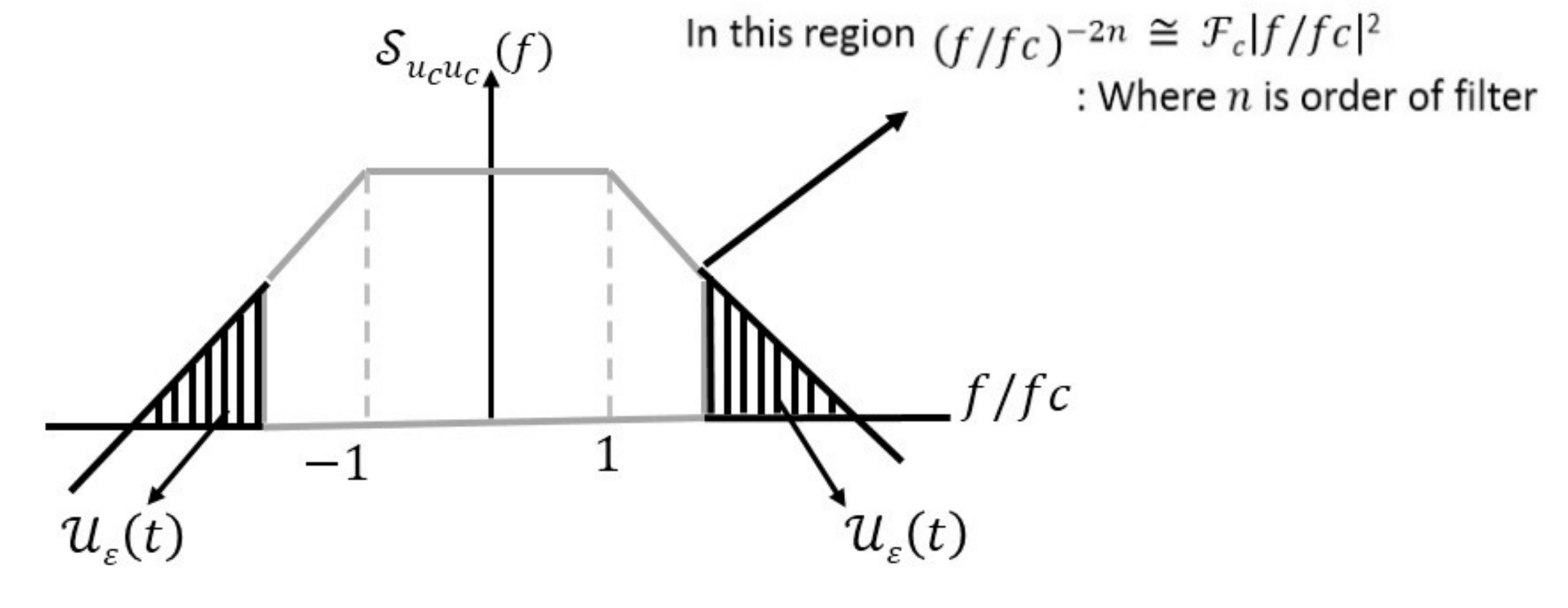}
\caption{Error in an actual signal}
\label{BLAnalysis}
\end{figure}

From the literature it is a well known fact that gain of the low-pass filter in the roll-off region varies as a factor of $$\left(\frac{f}{f_c}\right)^{-n} \approx \left| F_c \left(\frac{f}{f_c}\right)\right|,$$ where $f$ is the frequency of the signal, $f_c$ is the cut-off frequency, and $n$ the filter order [see Fig.\ref{BLAnalysis}]. The filter roll-off beyond the cut-off frequency is usually defined in dB/decade.

\paragraph{\textbf{Filtered White Noise}}The Power ($Pu_{\epsilon}u_{\epsilon}$) contained in the unexplained part of a band-limited signal generated by filtering white noise signal can be calculated by integrating the signal over the frequency band $[{f_s-f_c} ,\infty[$ i.e.

\begin{align}
Pu_{\epsilon}u_{\epsilon} &\cong 2\int_{f_s-f_c}^\infty \left(\frac{f_c}{f}\right)^{2n} df\notag\\
&\cong  \frac{2 f_c}{2n-1}  \left(\frac{f_c}{f_s-f_c}\right)^{2n-1}\notag\\
&\cong \frac{2 f_c}{2n-1}  \left(\frac{f_c}{f_s}\right)^{2n-1}
\label{eqn:PowerUe}
\end{align} because ${f_s-f_c} \cong f_s $ as $fs \gg f_c$. From (\ref{eqn:PowerUe}) it can be concluded that the power in the unexplained part of the band-limited signal $u_c(t)$ varies as 
\begin{equation}
Pu_{\epsilon}u_{\epsilon}=\MakeUppercase{\mathcal{O}}\left(\frac{f_c}{f_s} \right)^{{2n-1}}
\label{eqn:BLZOH}
\end{equation}

\begin{figure}[!h]
\centering
\captionsetup{justification=centering}
 \includegraphics[width=0.5\textwidth]{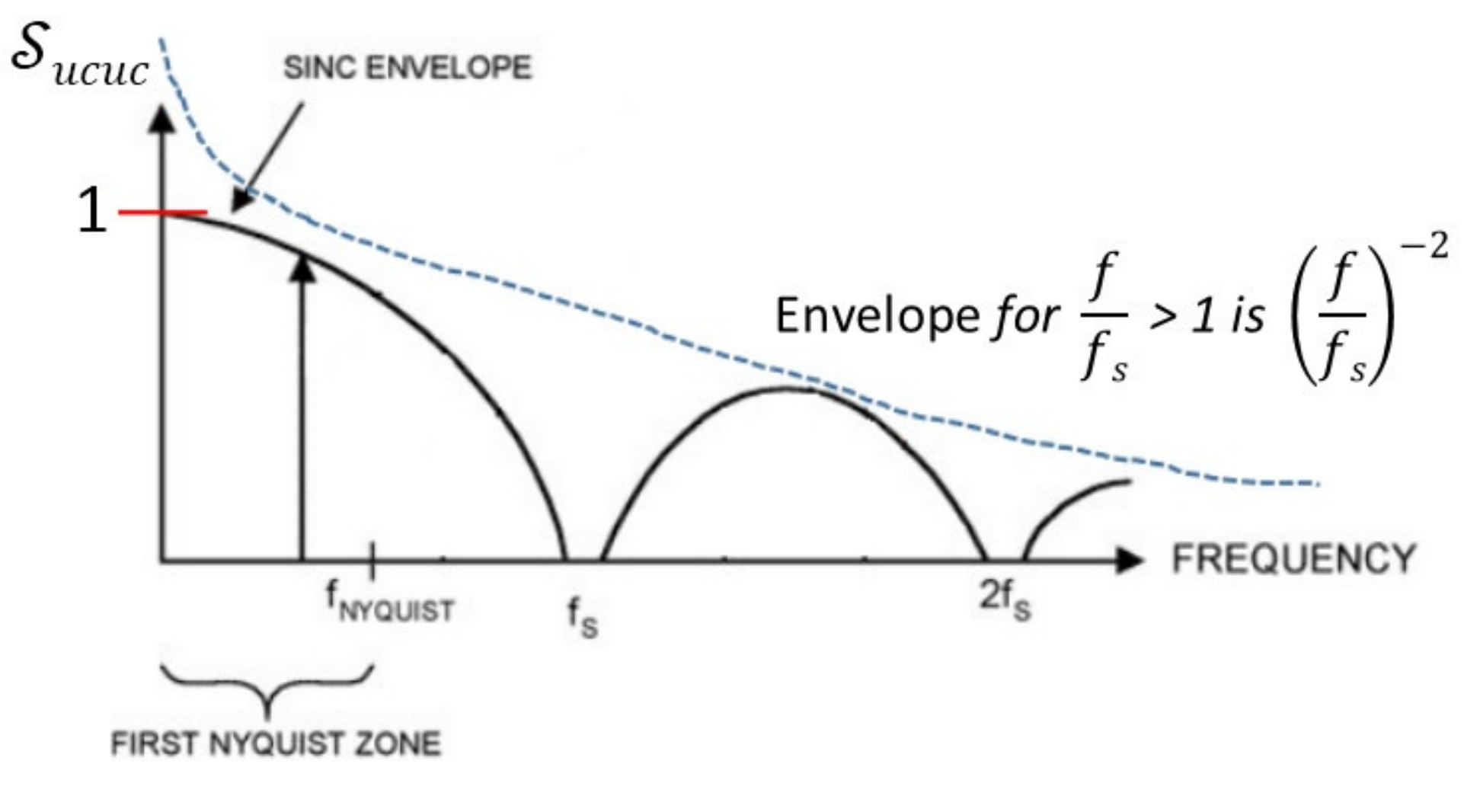}
\caption{ZOH Spectrum}
\label{ZOHSPEC}
\end{figure}

\paragraph{\textbf{ZOH White Noise}} In the case of the zero-order-hold (ZOH), the excitation signal is considered to be constant between consecutive samples. As seen from the envelope in the Fig.\ref{ZOHSPEC} \cite{RikJohan2012} , the envelope of $S_{u_cu_c}$ for $ \tfrac{f}{f_s} > 1$  is $(\tfrac{f}{f_s})^{-2}$,  hence the zero-order hold (ZOH) will create an additional roll-off and therefore it will not increase the order of magnitude of the error given in (\ref{eqn:BLZOH}) .

\paragraph{\textbf{Conclusion}}Therefore from the analysis above it can be concluded that

  \begin{equation}
 Pu_{\epsilon}u_{\epsilon}  = \left[ \left| u(t)-\hat{u(t)} \right|^2\right] \leq \MakeUppercase{\mathcal{O}} \left(\frac{f_c}{f_s} \right)^{{2n-1}}
\end{equation}for an all pole generator filter/actuator, independent of the ZOH or BL measurement of the signal.

\subsubsection{Error quantification of the output of a linear system}
The next step in the analysis of error is to observe the impact of the error in the one-step ahead prediction of $u(t)$ on the final output $y(t)$ of the discrete-time model. The next section provides a concise theoretical explanation of the impact of the error in $\hat{u}(t)$ on the final output $\hat{y}(t)$. For the purpose of quantifying the impact of the error in $\hat{u}(t)$ on the final output $\hat{y}(t)$ the following assumptions are made:


\begin{definition}
The data can be acquired at sufficiently high sampling rates and effect of sampling zeros, folding, etc. can be neglected. 
\end{definition}

\begin{figure}[!h]
\centering
\captionsetup{justification=centering}
 \includegraphics[width=0.35\textwidth]{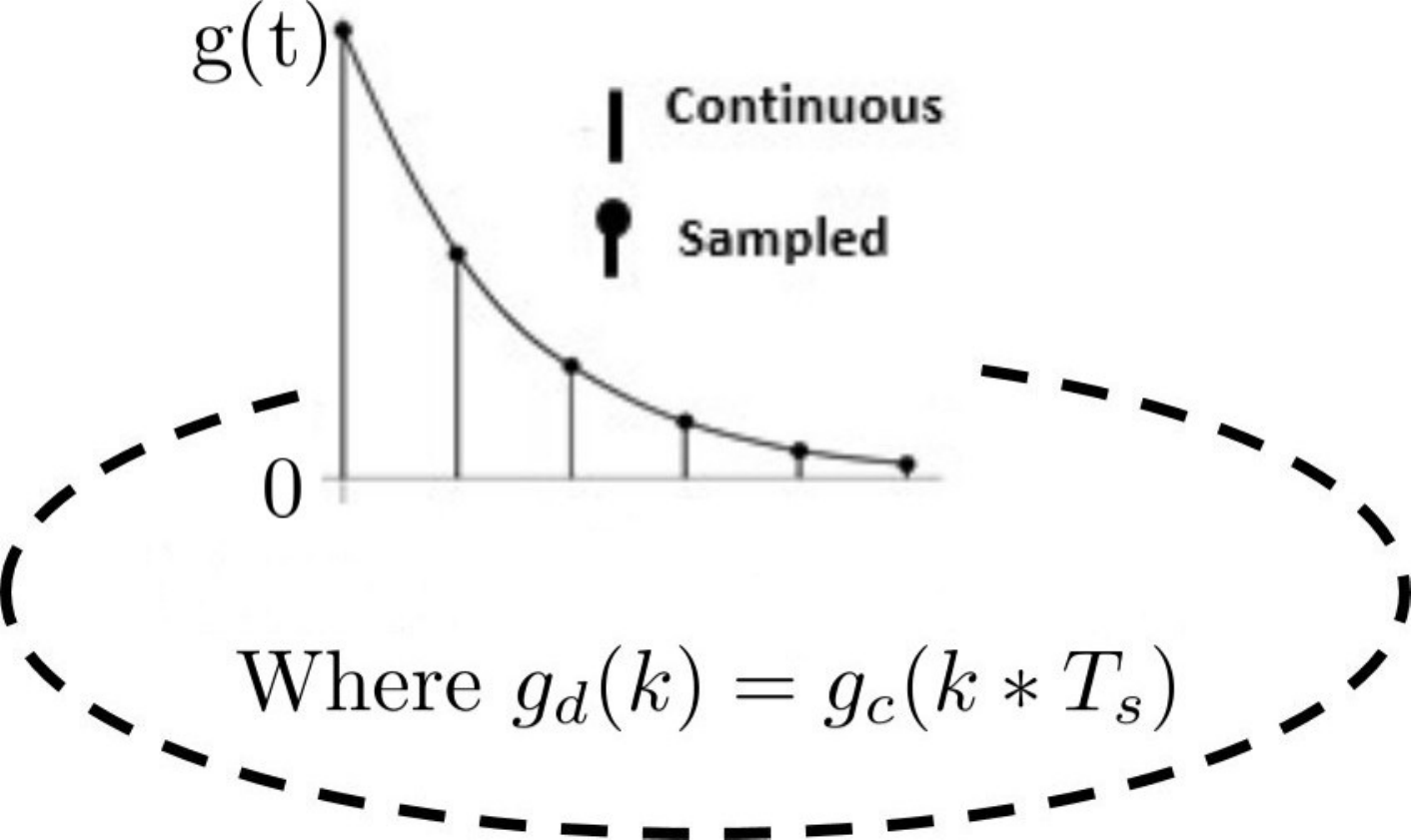}
\caption{Impulse response  for system with rel. degree $= 1$}
\label{IITCS}
\end{figure}

\begin{figure}[!h]
\centering
\captionsetup{justification=centering}
 \includegraphics[width=0.4\textwidth]{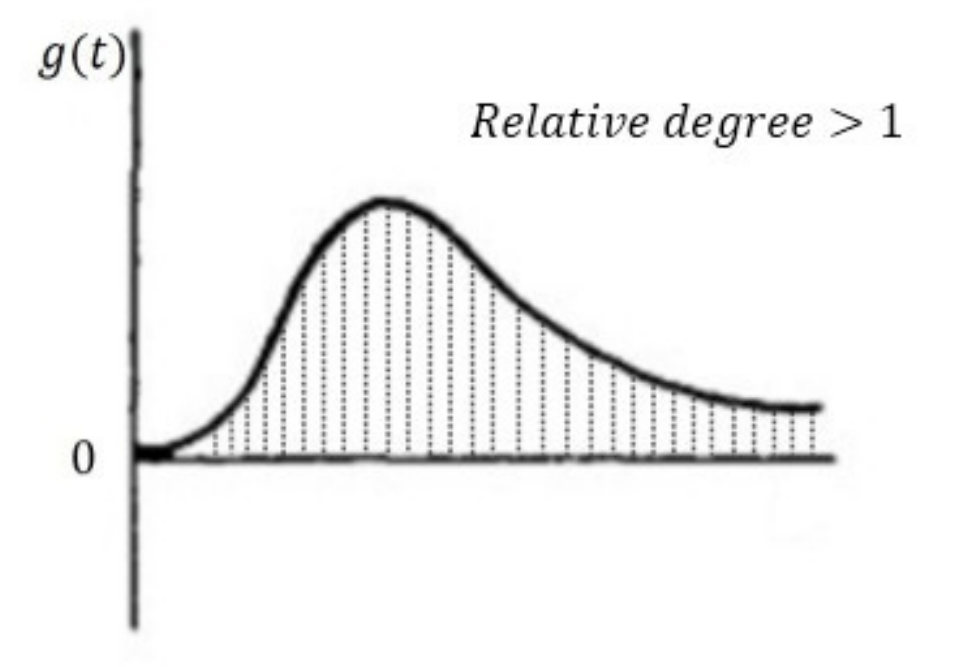}
\caption{Impulse response for system with rel. degree $> 1$ }
\label{IITRel2}
\end{figure}

\textbf{Remark}: This can also be resolved using virtually upsampling the data.
\begin{definition}
The discrete-time model representation of the continuous-time system with any arbitrary relative degree equal to \textit{d} will be close to the impulse invariant transform (I.I.T) \cite{kuo1966system,oppenheim2013discrete, papoulis1962fourier,Mecklenbrauker2000,Jackson2000} of the continuous-time impulse response as shown in Fig.\ref{IITCS} and Fig.\ref{IITRel2} respectively.
\end{definition}

\textbf{Remark}: The theoretical analysis is based on the discrete-time impulse response representation of the continuous time system, but it is equally valid for any other discrete-time model representation.

\subsection{Impact Of The Error In $\hat{u}(t)$ On The Final Output $\hat{y}(t)$}
\label{OutErr}
Using  the  initial-value  theorem  of the Laplace transform, the impulse response $g_{c}(t)$ of the continuous time system with relative degree $d > 1$ \cite {Eitelberg2006, Nelatury2007,Swaroop2001} meets.

\begin{equation}
	 {g_c}^{d-1}(t)\Big|_{t=0}=0 
\label{eqn:ImpulseResponse}
\end{equation} The output of the identified discrete-time model can be expressed as:
\begin{align}
y(t)&= \sum\limits_{k=0}^{\infty}{g}_{d}(k)u_d(t-k)\notag\\
	  &\cong \sum\limits_{k=0}^{\infty}g_c(kT_{s})u_d(t-k) 
\label{eqn:Impulse}
\end{align} Where ${g}_{d}(k) = g_c(kT_{s})$ due to impulse invariant transformation. From (\ref{eqn:ImpulseResponse}) it follows that $g_d(0)$ will converge to zero if $d \geq 2$, for $f_s\to\infty $. In the rest of this paper, we assume that $|g_d(0)| < M$, where $M$ is a bounded value of the response and if $d\geq2$ then $\lim\limits_{f_s\to\infty} M = 0$, .

The output $y(t)$ can further be expanded as described by  (\ref{eqn:DiscreteModel}), where $\hat{u}(t)$ is the one-step ahead prediction of the input signal $u(t)$.

\begin{equation}
\hat{y}(t)=\frac{1}{f_{s}}{g}_{d}(0)\hat{u}(t)+\frac{1}{f_{s}}\sum\limits_{k=1}^{\infty}{g}_{d}(kT_{s})u(t-k)
\label{eqn:DiscreteModel}
\end{equation} From  (\ref{eqn:Impulse}) and (\ref{eqn:DiscreteModel}), the error in the output signal can be written as
\begin{align}
y_{\epsilon}(t)&= {y}(t)-\hat{y}(t)\notag\\
&= \frac{1}{f_{s}} g(0) u_{\epsilon}
\label{eqn:Error}
\end{align} Furthermore the power contained in the output error signal can be expressed as \begin{align}
 Py_{\epsilon}  &= \left(\frac{1}{f_{s}}\right)^2 g(0)^2 Pu_{\epsilon}\notag\\
&= \left(\frac{1}{f_{s}}\right)^2g(0)^2 \MakeUppercase{\mathcal{O}} \left(\frac{f_c}{f_s} \right)^{{2n-1}}
\label{eqn:ErrorY}
\end{align} This implies that the root mean squared error in ${y_{\epsilon}}_{RMS}$ is upper-bounded by
\begin{equation}
{y_{\epsilon}}_{RMS} \leq  \frac{1}{f_s}\MakeUppercase{\mathcal{O}} \left(\frac{f_c}{f_s} \right)^{{n-\frac{1}{2}}}
\label{eqn:ErrorUe}
\end{equation} (\ref{eqn:ErrorY}) and  (\ref{eqn:ErrorUe}) above describe a relationship between the cut-off frequency of the generator filter, the sampling frequency and the error in the output as well as the unmodelled part of the input signal respectively. In the next section a qualitative experimental investigation has been performed to validate this theoretical analysis.
\section{\textbf{Experimental Verification}}
In order to validate the theoretical results qualitatively, real-world experimental investigations were performed. In the sections below, first the measurement set-up is introduced, next the experiment design is explained, and finally the results are discussed. A preliminary findings of the results has been already presented in the \cite{Rishi2015I2MTC}

\subsection{Measurement Set-up}

\begin{figure}[!h]
 \centering
\captionsetup{justification=centering}
 \includegraphics[width=0.5\textwidth]{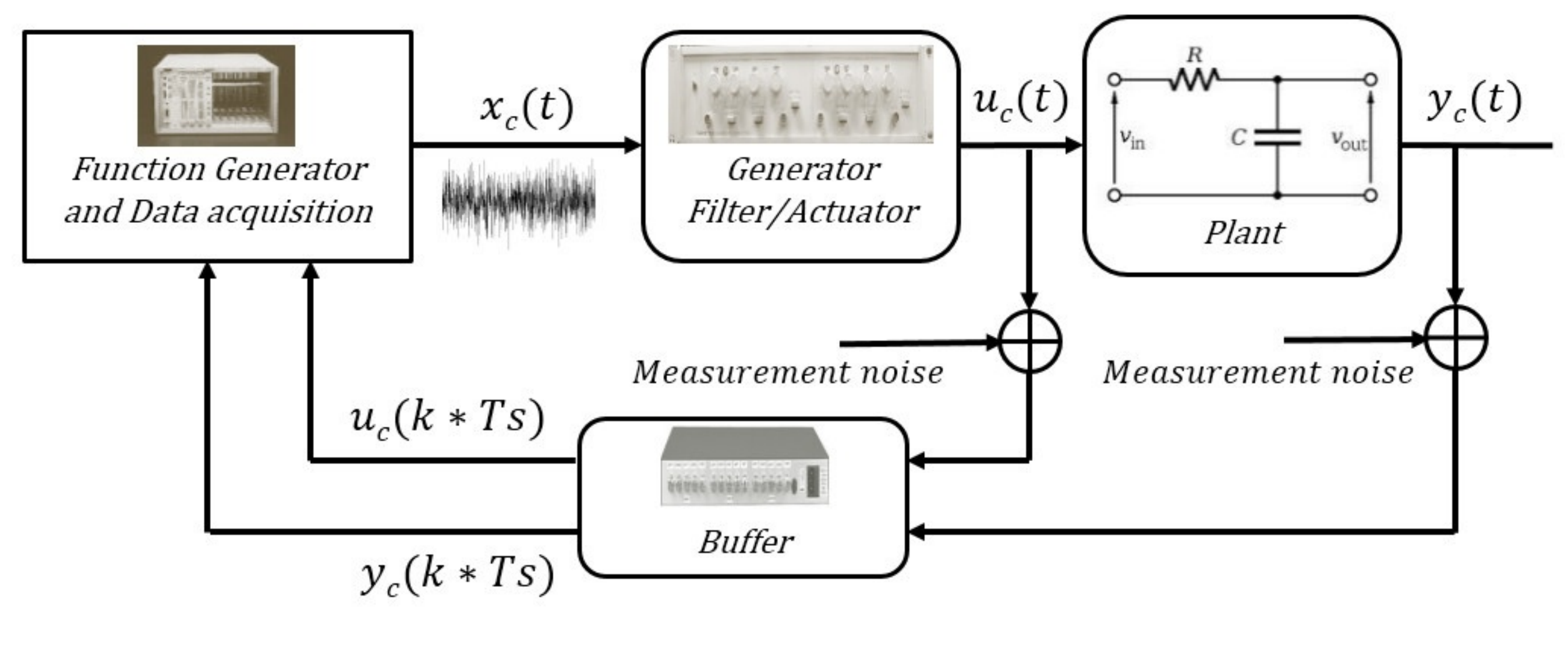}
\caption{Experimental set-up}
\label{Experiment}
\end{figure}

\subsubsection{\textbf{Linear System}}
Fig.\ref{Experiment} demonstrates the schematic of the experimental set-up  and the measurement architecture for this validation study. For the sake of simplicity a $R-C$ filter is selected as the continuous-time plant to be identified in the experiment involving the identification of a linear system. Since in this case the relative degree $d=1$, it is a worst case example because $g(0)$ will be the dominant term in the impulse response. In general, it can be any other real continuous-time system. $x_c(t)$ denotes the ideal reference signal from the function generator whereas $u_c(t)$ and $y_c(t)$ are the actual continuous time input and the output signal of the plant respectively .

\subsubsection{\textbf{Nonlinear system}}
In the case of a nonlinear system, the plant shown in Fig.\ref{Experiment} is replaced by the Silverbox while keeping the other experimental set-up/measurement methodology the same. The Silverbox is an electrical circuit,  simulating a mass-spring-damper system.  It is an example of nonlinear dynamic system with feedback as shown in Fig.\ref{ExperimentSilverbox}, where the linear contributions are dominant for the small excitation levels of the input signal \cite{Schoukens2003}. The system's behaviour can be approximately described by the following equation: 

\begin{figure}[!h]
 \centering
\captionsetup{justification=centering}
 \includegraphics[width=0.5\textwidth]{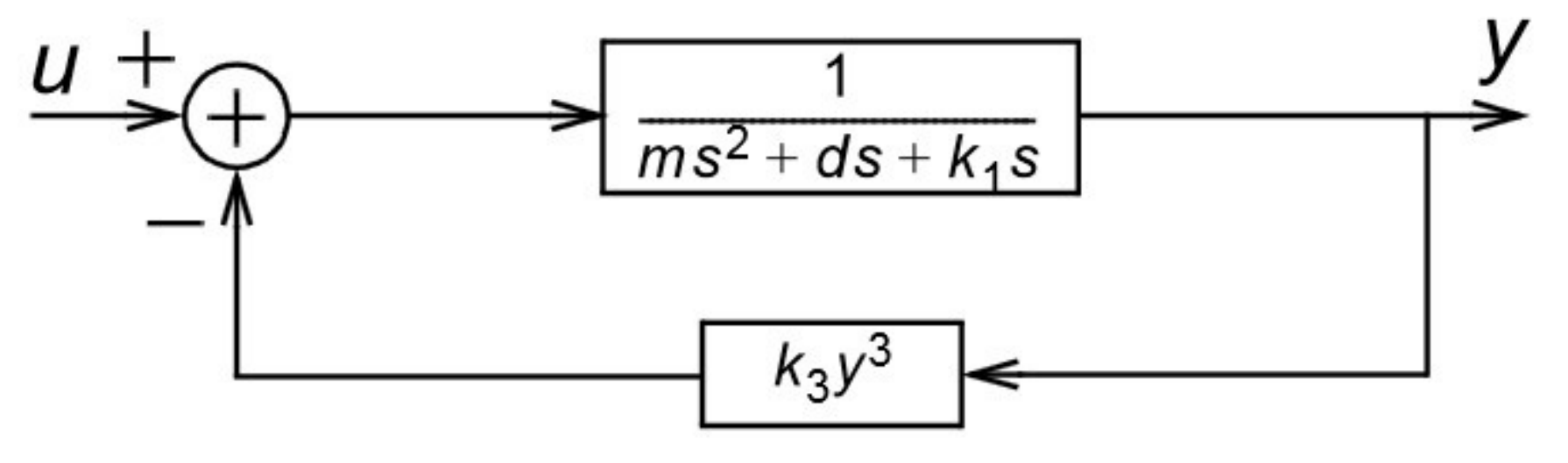}
\caption{Silverbox Dynamics}
\label{ExperimentSilverbox}
\end{figure}

\begin{equation}
m\ddot{y}(t)+ d\dot{y}(t)+k_1y(t)+k_3y^3(t) = u(t)
\label{eqn:Silverbox}
\end{equation} where $u(t)$ represents the input force applied to the mass $m$, and the output $y(t)$ is the mass displacement. Parameters $k_1$ and $k_3$ describe the (nonlinear) behavior of the spring, and $d$ is the damping of the system \cite {RikJohan2012}.

As shown in the measurement set-up in Fig.\ref{Experiment}, the signals are generated by an arbitrary waveform  generator (AWG) or function generator, the Agilent/HP E1445A, with an internal reconstruction filter that has a  cut-off frequency at 250 kHz. The output of the generator filter is filtered by a $4^{th}-order$ Wavetek Dual Hi/low pass (Model 432) filter with a cut-off frequency of 100 Hz. The input and output signals of the plant (analog RC Filter with a cut-off frequency of 1kHz / Silverbox) are measured by the alias protected acquisition  channels (Agilent/HP  E1430A). The AWG and acquisition cards are clocked by the AWG clock, and hence the acquisition is phase coherent to the AWG. Finally,  buffers  are  added  between  the  acquisition  cards and  the  input  and  output  of  the  device under test (DUT)  to  impose  impedance isolation of the signals. The buffers are added to match the 50 $\Omega$
input impedance of the Agilent/HP E1430A VXI modules acquisition channels to a high impedance input. The buffers are very linear ( $\approx$ 85 dBc at full scale and 1 MHz) up to 10 V peak to peak, and have an input impedance of 1 M$\Omega$ and a 50 $\Omega$ output impedance. 

\subsection{Experiment Design}
A normally distributed noise signal (white noise) is used as an input excitation signal for the identification of the linear model, and for the identification of the polynomial nonlinear state-space (PNLSS) model. An odd-random phase multisine signal \cite{RikJohan2012} is used  to excite the Silverbox in the frequency band of [0-100 Hz]. In an odd-random phase multisine signal, only the odd frequency lines are excited with the user-defined amplitude levels. The even frequency lines as well as the non-excited odd lines, then act as the detection lines for the detection of system nonlinearities. The choice of the input excitation signal is not only restricted to these two classes of signals, rather one can also use any persistently exciting signal such as e.g. a flat spectrum random phase multisine or a uniformly distributed noise signal as an input excitation signal.  We make this choice in order to verify the level of the nonlinear distortions during the experiment. The excitation signal has a period of 78125 samples. The level of the input excitation is zero mean with a standard deviation 0.99 V for the first experiment involving the identification of the linear model and the amplitude of the full odd random phase multisine is zero mean with a standard deviation of 127mV during the identification of the polynomial nonlinear state-space (PNLSS) model. The three different experiments performed in this investigation are:
\begin{enumerate}
 \item \textbf{The one-step ahead prediction of the generator/actuator signal $u(t)$:} For the one-step ahead prediction the data is acquired for $P = 1$ period at different bandwidths of the generator filter/actuator while keeping the sampling frequency $f_s$ constant at 78.125 kHz. For the sake of brevity an Auto-regressive with an exogenous input (ARX) model structure is chosen for the one-step ahead prediction.  The one-step ahead prediction is performed at different bandwidths of the generator filter as well as for different model orders.
 \item \textbf{Identification of a discrete-time model  based on the sampled $u_c(t)$ and $y_c(t)$:} $P = 2$ periods of data is acquired at a sampling frequency $f_s$ of 156.25 kHz for the model identification experiment at a constant generator filter/actuator bandwidth. The data is down-sampled virtually for identifying discrete-time models at the different sampling frequencies. For the identification of the discrete-time model, the Output-error (OE) model structure ($y(t)= \frac{B(q)}{F(q)}u(t-nk)+e(t)$) is chosen for the same reasons. During the identification of the model with $B_0=0$, the complexity (order) of the model was slightly increased (numerator $n_b=2$, denominator $n_f =4$ from numerator $n_b=2$,  denominator $n_f=2$ with $B_0$  term intact) to accommodate the extra dynamics.
 \item \textbf{Identification of a polynomial nonlinear state-space (PNLSS) discrete-time model  based on the sampled $u_c(t)$ and $y_c(t)$:}. In order to identify the discrete-time nonlinear model for the silverbox, the polynomial nonlinear state-space model structure \cite{Paduart2010}  is selected because it implicitly embeds the delay in its model structure as described below:
 \begin{align}
 x(t+1) &= Ax(t) + Bu(t) + E\zeta(t)\\
 y(t) &= Cx(t) + Du(t) + F\eta(t)
\label{eqn:PLNSS}
\end{align}

The coefficients of the linear terms in $x(t)$  and $u(t)$  are given by the matrices $A \in \mathbb{R}^{n_a \times n_a}$ and $B \in \mathbb{R}^{n_a \times n_u}$ in the state equation,  $C \in \mathbb{R}^{n_y \times n_a}$ and $D \in \mathbb{R}^{n_u \times n_a}$ in the output equation. The  vectors $\zeta(t) \in \mathbb{R}^{n_{\zeta}}$ and $\eta(t) \in \mathbb{R}^{n_{\eta}}$ contain nonlinear monomials in $x(t)$ and $u(t)$ of  degree  two  up  to  a  chosen  degree  $P$ .  The coefficients associated with these nonlinear terms are given by the matrices $E \in \mathbb{R}^{n_a \times n_{\zeta}}$ and $D \in \mathbb{R}^{n_y \times n_{\eta}}$. Note that the monomials of degree one are included in the linear part of the PNLSS model structure. $P = 2$ periods of data for $M=10$ different realisations of the odd random phase multisine input excitation is acquired at a sampling frequency $f_s$ of 78.125 kHz for the model identification experiment at a constant generator filter/actuator bandwidth. $M = 9$ realisations are used for the training purpose whereas $M = 1$ realisation is kept aside to test the model performance on an unseen validation data. For the analysis purpose, the sampled data was virtually down-sampled at different sampling rates by explicitly omitting the samples.

\end{enumerate} 
\subsection{Results}

\begin{figure}[!h]
 \includegraphics[width=0.5\textwidth]{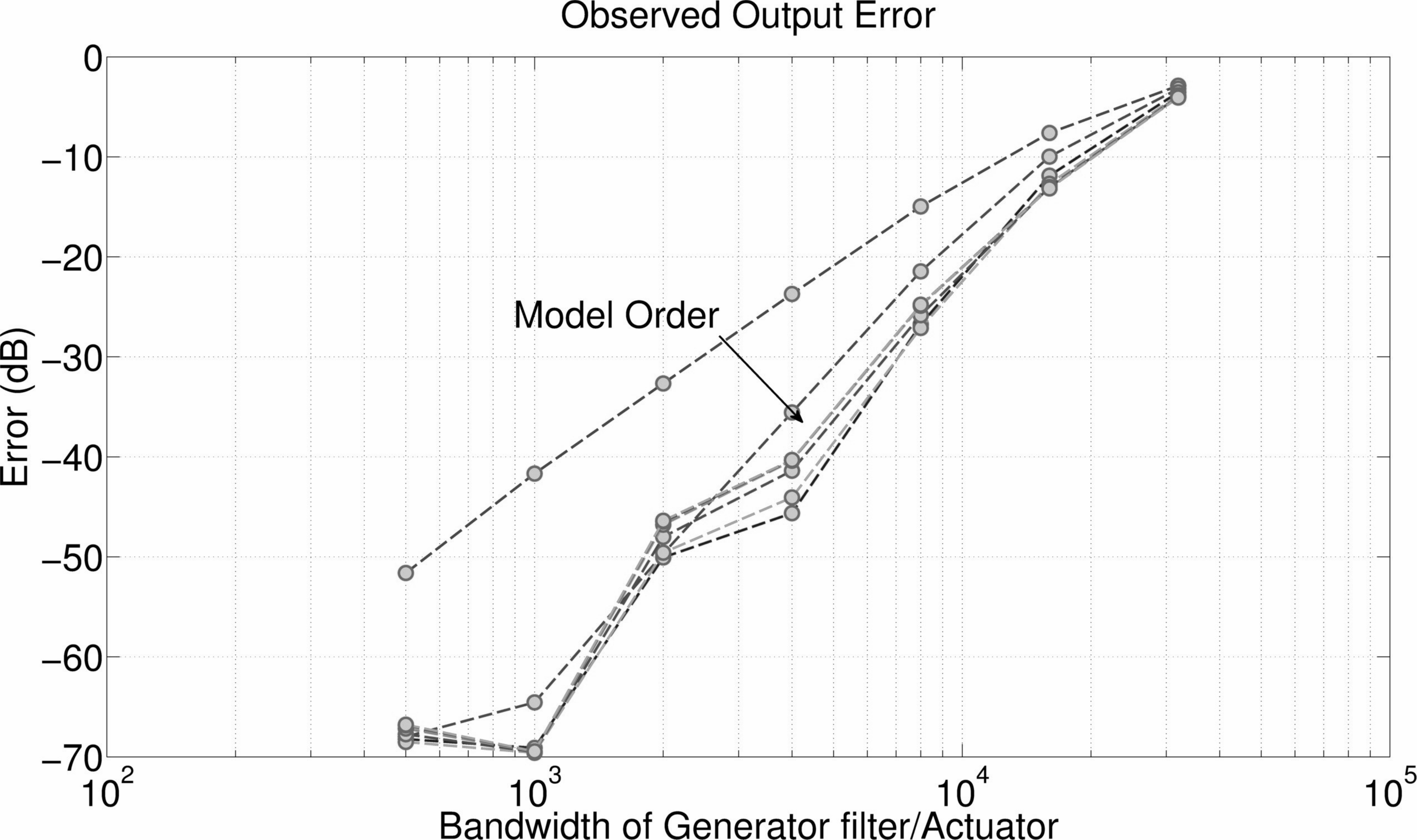}
\caption{The one-step ahead prediction error of the ARX model structure at different bandwidths of the generator filter and at a fixed sampling frequency $f_s = 78.125$ kHz.}
\label{Experiment2}
\end{figure}

Fig.\ref{Experiment2} shows the evolution of the Root Mean Square Error (RMSE) in the one-step ahead prediction of an actual band-limited signal for different model orders against the bandwidth of the generator filter/actuator. It can be clearly observed that the RMSE of the one-step ahead prediction varies as a function of the generator filter bandwidth and ultimately converges to a constant (Maximum) value. This supports the argument made in the section \ref{OneStepTheory}. 

\subsubsection{\textbf{Linear System}}

\begin{figure}[!h]
 \centering
 \includegraphics[width=0.5\textwidth]{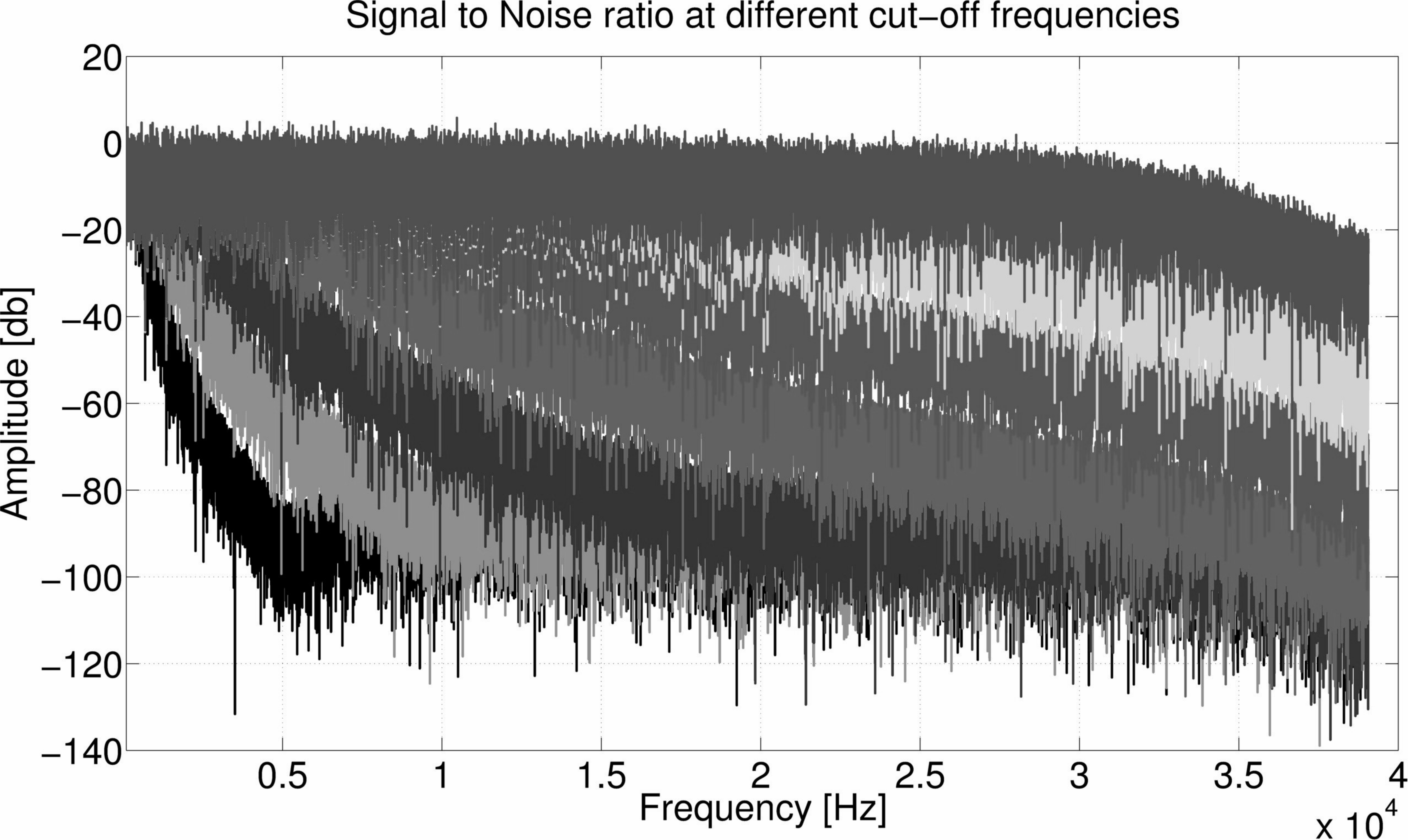}
\caption{The power spectrum of the measured $u_{c}$ at the different bandwidths of the generator filter for a fixed sampling frequency $f_s = 78.125$ kHz. It is evident from the figure that the signal around the cut-off frequency ($1$kHz) of RC filter is not persistently exciting (due to poor SNR) for the identification of the linear model}
\label{Experiment3}
\end{figure}

Fig.\ref{Experiment3} provides an explanation to the slight dip observed in the RMSE of the one-step ahead prediction for low model orders at around $1$ kHz  bandwidth of the generator filter. It shows the spectral analysis performed using the Hanning window on the signals acquired at different bandwidths of generator filter. It clearly shows that at the lower bandwidths of the generator filter, the signal-to-noise (SNR) is much lower than at the higher bandwidths. 

\begin{figure}[!h]
 \centering
 \includegraphics[width=0.5\textwidth]{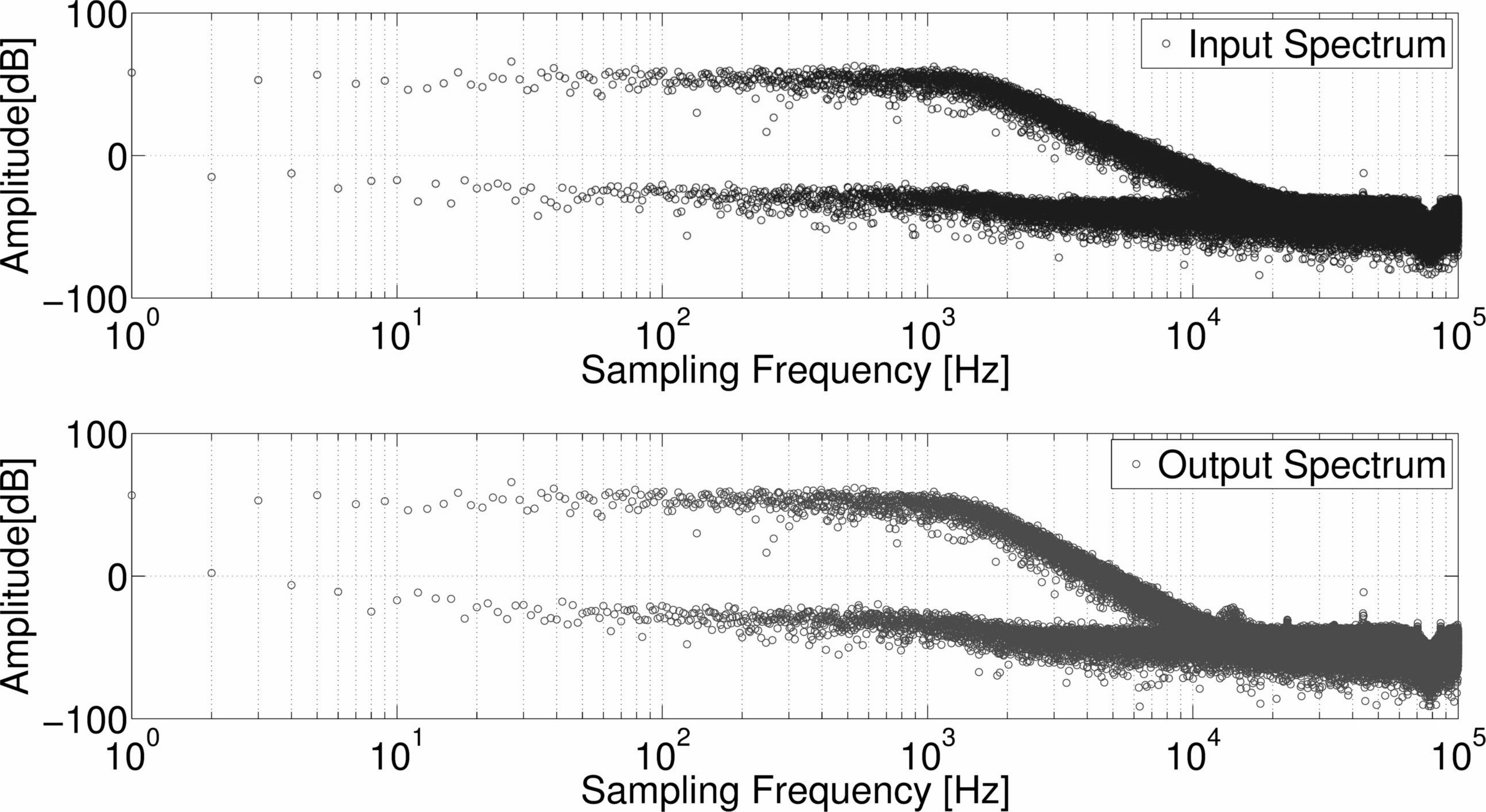}
\caption{Spectrum of the input and output signals of the linear system}
\label{Experiment5}
\end{figure}

\begin{figure}[!h]
 \centering
\captionsetup{justification=centering}
 \includegraphics[width=0.5\textwidth]{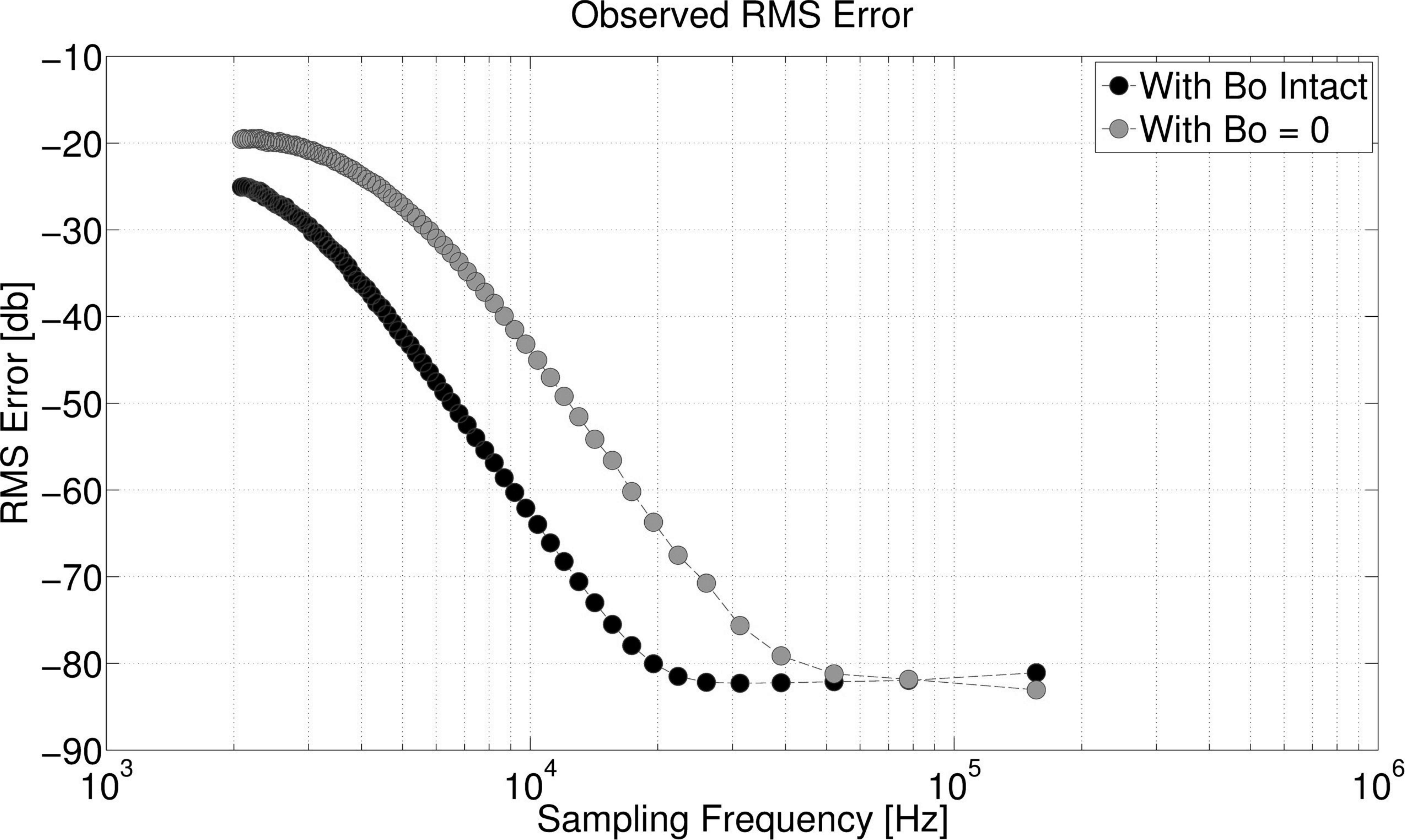}
\caption{Output error of the linear model}
\label{Experiment4}
\end{figure}

The Fig.\ref{Experiment5}, shows the frequency spectrum of the input and output signals of the linear system and the results obtained from the identification of the discrete-time model for the continuous time first order linear dynamical system with and without forcing the direct term $g_d(0)=0$ are shown in Fig.\ref{Experiment4}. It is clearly observed that the influence of explicitly forcing the direct-term $g_d(0)=0$ diminishes very quickly, if the data is acquired at sufficiently high sampling frequencies. It is clearly seen that the slope of the error curve with $B_{0}=0$ is approximately $\approx -75dB/decade$. The theoretical prediction made in Section  \ref{OutErr}, corresponds to $\approx -90dB/decade$ because the $1^{st}-order$ linear system was excited by the white Gaussian noise filtered with a $4^{th}-order$ low-pass filter. The observed drop in error is slightly less than  as per the theoretical prediction. The reason behind this can easily be understood by carefully looking at the Fig.\ref{Experiment5}, which clearly shows that the input excitation rolls-off very slowly above $10kHz$. This is a voilation of the low-pass assumption that is made in the developed theroy, hence along with the presence of the measurement noise, it hinders the achievement of the error bound predicted by the theory exactly.

The RMSE of the discrete-time model with forced delay term reduces very quickly afterwards with respect to the sampling frequency and ultimately converges to the same minimum value (lower bound) which is observed while keeping the direct term intact during the identification of discrete-time model of a particular pre-specified model order as well as model structure.  

\subsubsection{\textbf{Non-linear System}}
The results obtained from the identification of the polynomial nonlinear state-space (PNLSS) model is shown in Fig.\ref{PNLSSERRROR}. As discussed earlier, this particular discrete-time model structure implicitly embeds the forced delay term.   

\begin{figure}[!h]
 \centering
 \includegraphics[width=0.5\textwidth]{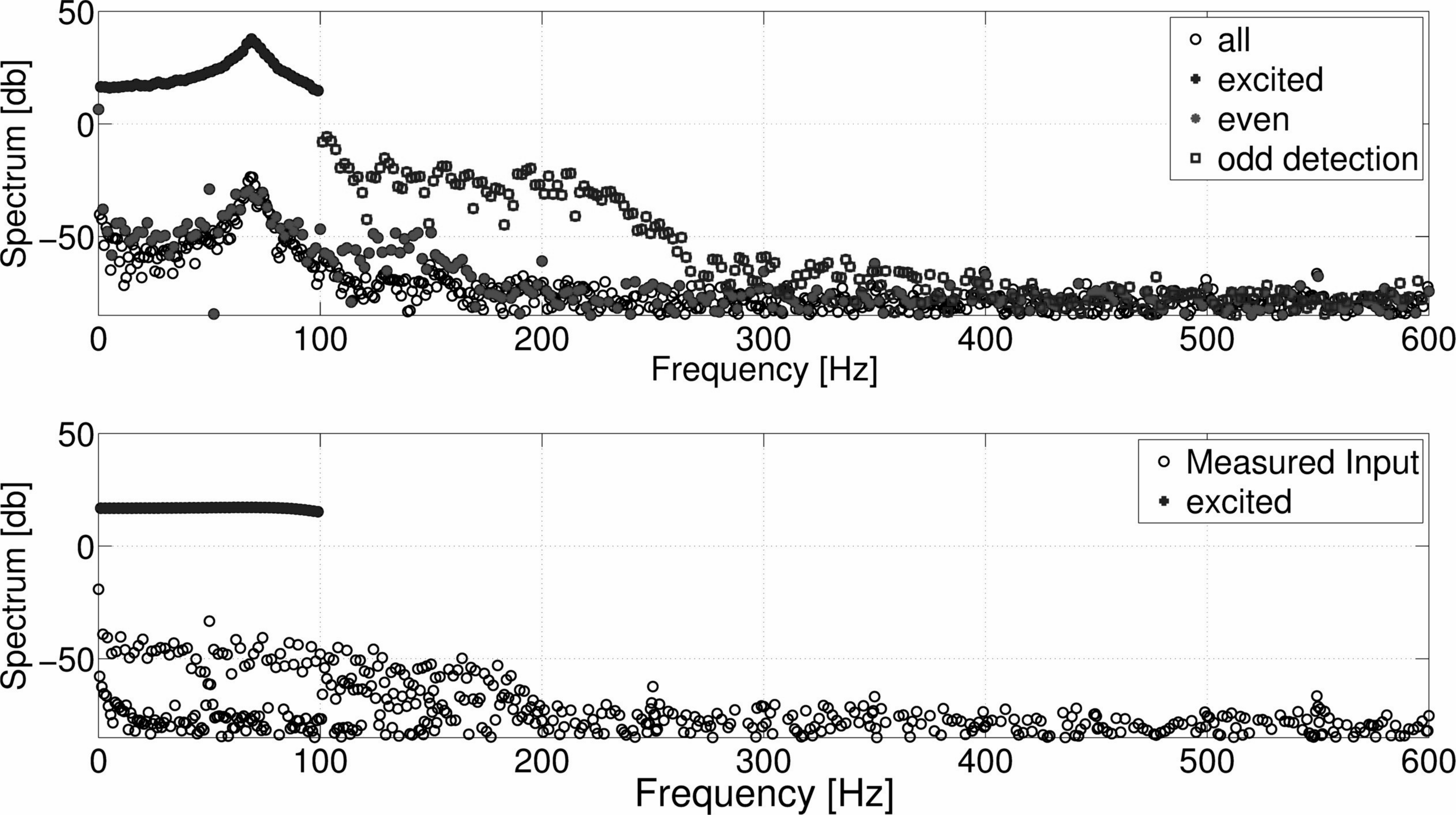}
\caption{The input and output spectrum of the nonlinear silverbox system}
\label{OutPutSpecPNLSS}
\end{figure} 

Fig.\ref{OutPutSpecPNLSS} shows the input and output spectrum of the silverbox dynamics excited by an odd random phase multisine input signal. From the output spectrum, it can be observed that the first resonance peak of the system lies at around $\approx 70 Hz$. In order to completely capture the information about this resonance peak inside the PNLSS discrete-time model structure, we must at least sample the input and output signals at atleast $\approx 140 Hz$ or at a greater sampling frequency. 

{Fig.\ref{PNLSSERRROR}, shows two metrics namely $y_{rms}$ and $y_{relative}$ for the output error of the PNLSS model, which are defined below. The mean value of all the signals is removed in order to eliminate the effect of the offset error, that might be present in the measurement setup.

\begin{align}
 y_{rms} &= \sqrt{ \frac{\sum\limits_{n = 1}^{n} {(\tilde{y}_{val} - \tilde{y}_{mod})}^2}{n}}\\
 y_{relative} &= \frac{\sqrt{ \frac{\sum\limits_{n = 1}^{n} {(\tilde{y}_{val} - \tilde{y}_{mod})}^2}{n}}}{\sqrt{\frac{\sum\limits_{n = 1}^{n} {({\tilde{y}_{val})}^2}}{n}}},
\label{eqn:Metrics}
\end{align} where $\tilde{y}_{val} = y_{val} - \mu_{val}$, $\tilde{y}_{mod} = y_{mod} - \mu_{mod}$,  $n$ is the number of data samples, $\mu$ is the mean, $y_{val}$ is the measured output and $y_{mod}$ is the model output respectively. From the Fig.\ref{PNLSSERRROR}, it is clearly observed that, the RMSE of PNLSS discrete-time model structure for a sampling frequency of $200Hz$ is $\approx -25dB$ which can be further reduced to a level of $\approx -44dB$ just by doubling the sampling frequency. It can also be seen that the error diminishes very quickly with respect to the sampling frequency for both the training and validation data sets.

\begin{figure}[!h]
 \centering
\captionsetup{justification=centering}
 \includegraphics[width=0.5\textwidth, height=0.25\textheight]{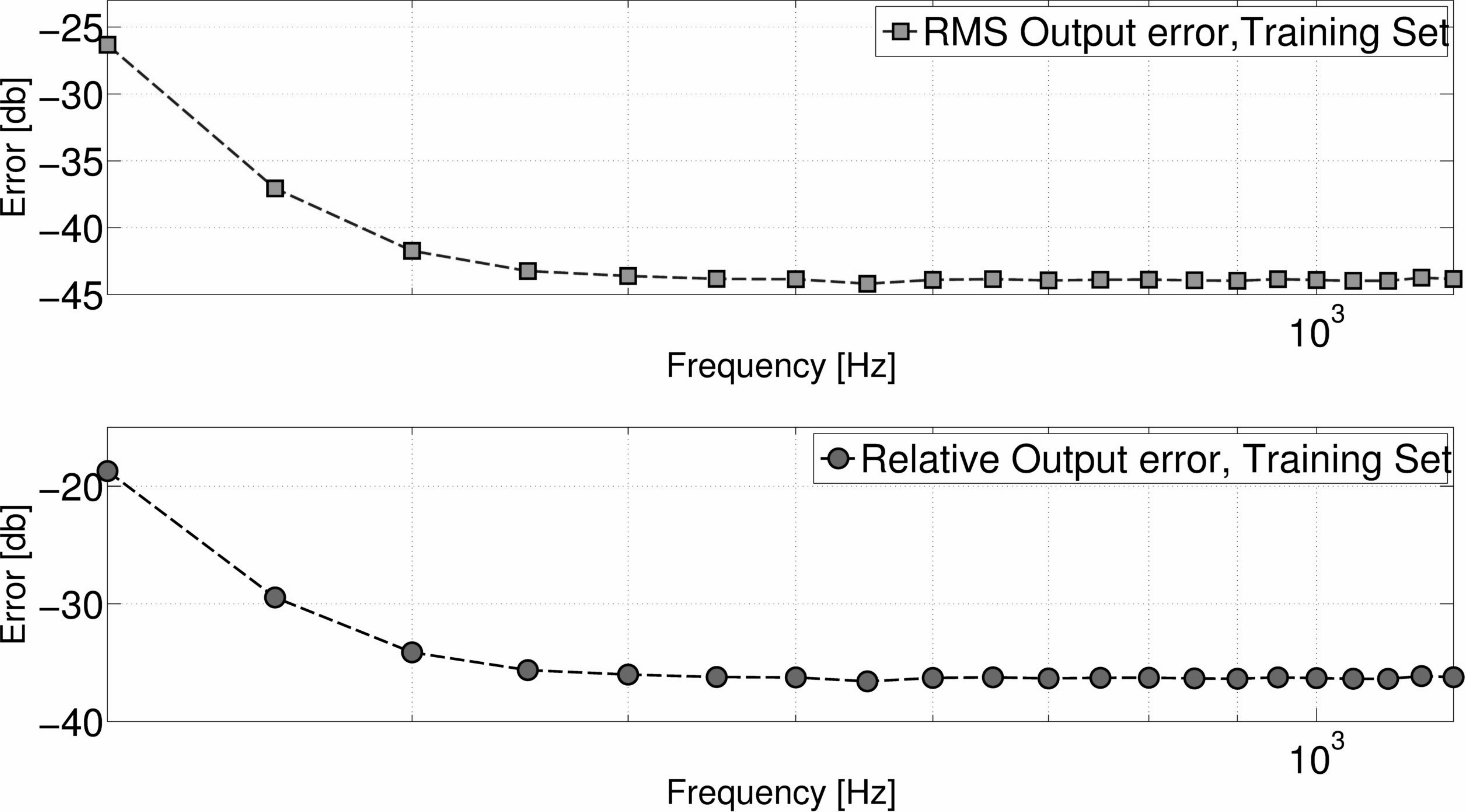}
\caption{Output error (Training Set) of PNLSS Model}
\label{PNLSSERRROR}
\end{figure}

Additionally, from the Fig.\ref{PNLSSERRRORAliasing}, it is clear that the magnitude of the errors due to aliasing (which might be introduced due to explicitly omitting the data samples) is way below the discrete-time model errors. Fig.\ref{ErrorPNLSS} shows the variation of the spectrum of the model error for different sampling frequencies.

\begin{figure}[!h]
 \centering
\captionsetup{justification=centering}
 \includegraphics[width=0.5\textwidth, height=0.25\textheight]{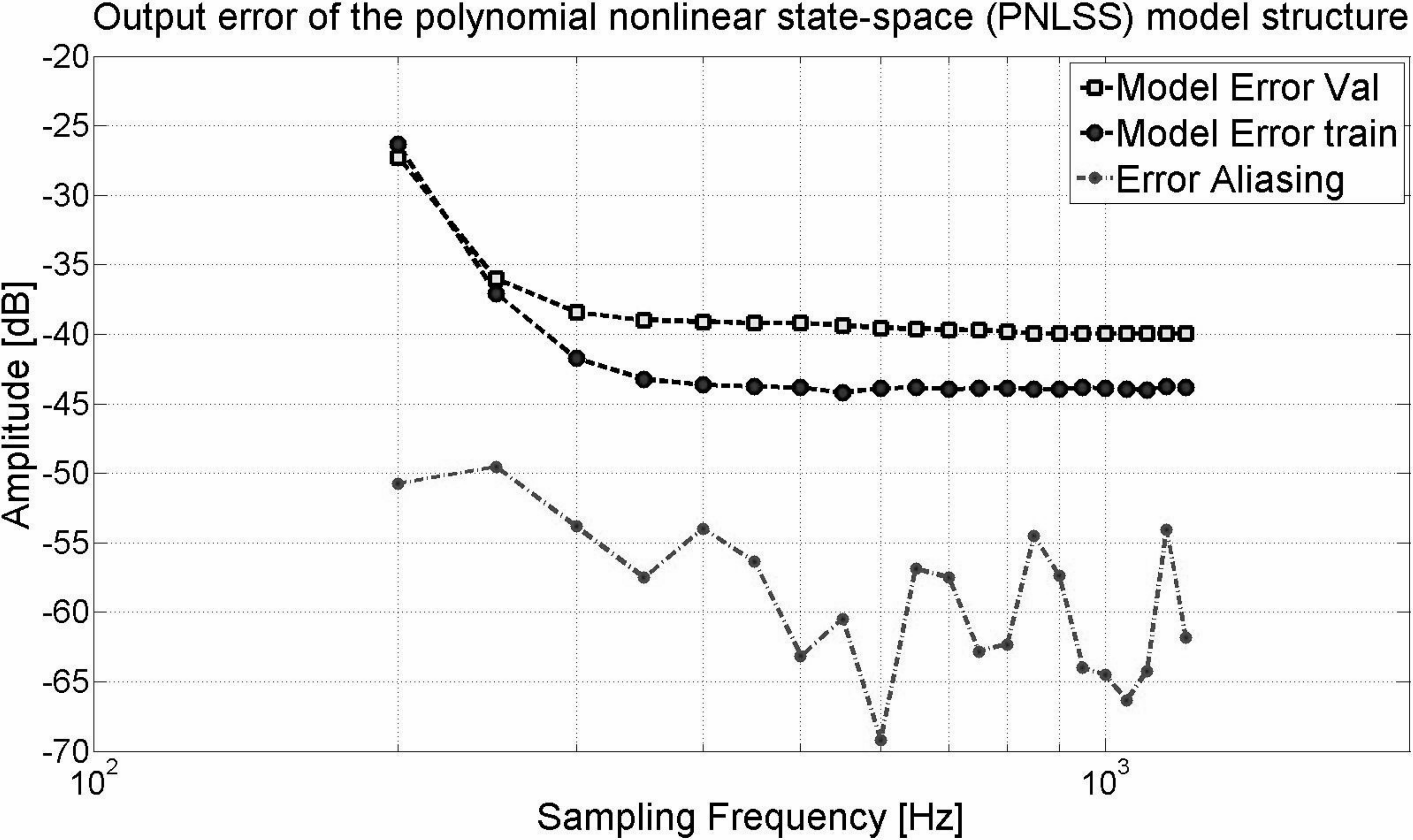}
\caption{Effect of Aliasing}
\label{PNLSSERRRORAliasing}
\end{figure} 

\begin{figure}[!h]
 \centering
\captionsetup{justification=centering}
 \includegraphics[width=0.5\textwidth, height=0.28\textheight]{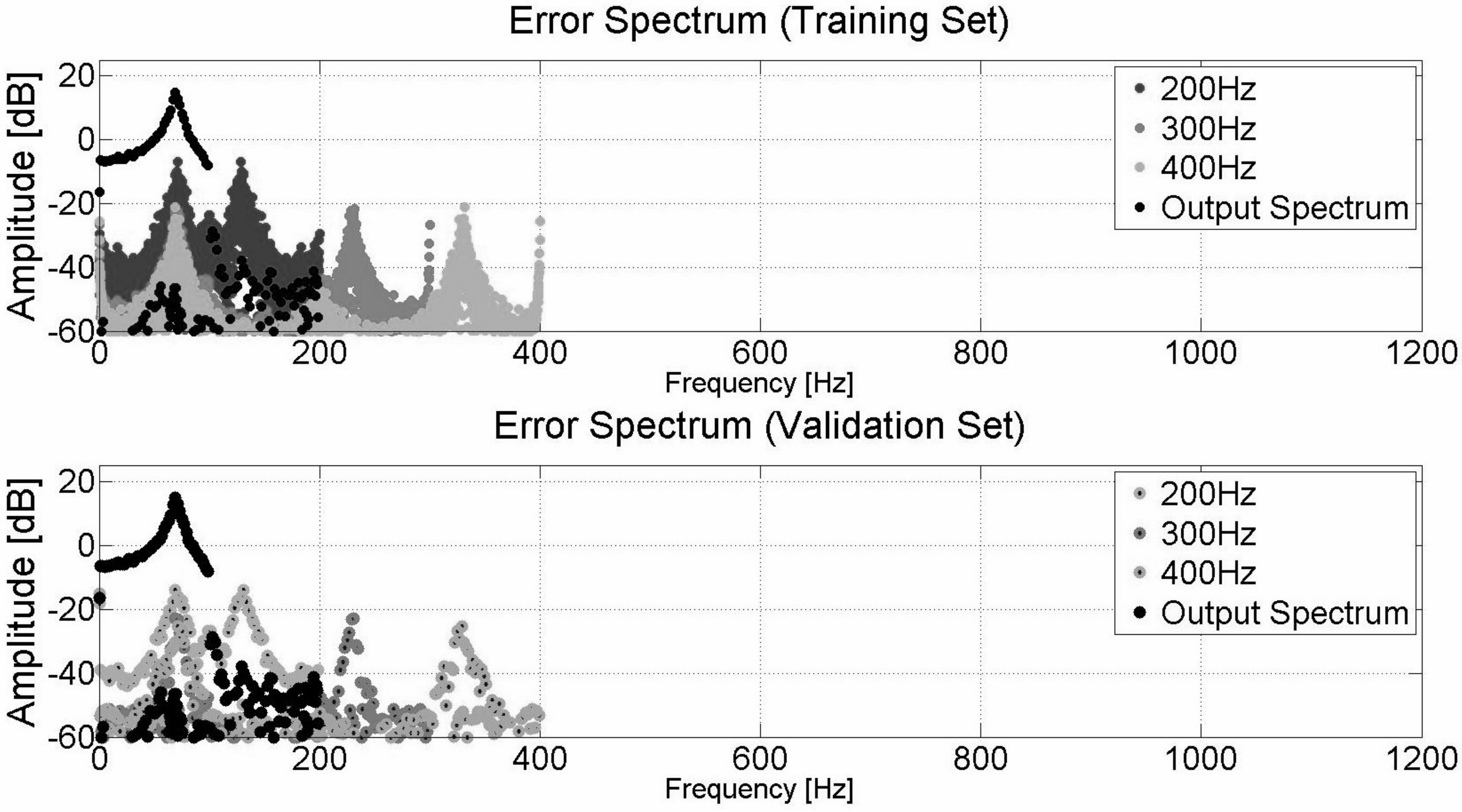}
\caption{Spectrum of PNLSS model error}
\label{ErrorPNLSS}
\end{figure} 


\section{\textbf{Conclusion}}
 
 For developing more realistic discrete-time models, often one has to work under band-limited assumptions as the ZOH assumptions does not hold at the output of the actuator that is driving the system. In this paper, we proposed a measurement approach for developing recursive discrete-time simulation models with direct-term $g_d(0)$ equal to 0 under the band-limited measurement assumptions. A theoretical expression involving the factors affecting the error bounds associated with these kind of linear models was also derived. 
Results obtained from the experiments were found consistent with the theory developed. These results also qualitatively support the theoretical reasoning provided in the Section \ref{Theory} which in-turn further extends our knowledge of the errors associated with the discrete-time models with forced delay under band-limited assumptions. 

The theoretical analysis and experimental investigation (both for linear as well as nonlinear system) reveals that, in order to develop good recursive discrete-time models with quantified error bounds, it is important to choose a good generator filter and explicitly introduce it before the continuous-time system to be identified. The sampling rates should be chosen adequately fast, e.g. at least 10 times the cut-off frequency of the generator filter. 

A sufficiently accurate model can also be obtained by up-sampling the data virtually, even if the data-acquisition set-up does not allow for very high sampling rates. Furthermore, the order of the identified discrete-time model can be increased to compensate for the error introduced by explicitly forcing the $g_d(0)$ equal to zero. 

This measurement approach, as well as the theoretical reasoning, is quite generic and can easily be applied to a wide class of dynamical systems including non-linear systems e.g. nonlinear feedback systems. The main advantage of the proposed methodology is that, good fast recursive discrete-time models with bounded output errors under band-limited measurement assumptions can be developed which eliminates the need to solve explicitly the nonlinear algebraic loops at each time step.

\bibliographystyle{IEEEtran}
\bibliography{References}

\end{document}